\documentclass[12pt,preprint]{emulateapj}
\usepackage{apjfonts,psfig}
\begin{document}
\title{\hspace{20pt}Surface Brightness Profiles of Galactic Globular Clusters from Hubble Space Telescope Images} 
\author{Eva Noyola \and Karl Gebhardt} 
\affil{Astronomy Department, University of Texas at Austin, Austin, TX 78712}
\email{eva@astro.as.utexas.edu, gebhardt@astro.as.utexas.edu}

\begin{abstract} 

Hubble Space Telescope allows us to study the central surface
brightness profiles for globular clusters at unprecedented detail. We
have mined the $\it{HST}$ archives to obtain 38 WFPC2 images of
galactic globular clusters with adequate exposure times and filters,
which we use to measure their central structure. We outline a reliable
method to obtain surface brightness profiles from integrated light
that we test on an extensive set of simulated images. Most clusters
have central surface brightness about 0.5 mag brighter than previous
measurements made from ground-based data, with the largest differences
around 2 magnitudes. Including the uncertainties in the slope
estimates, the surface brightness slope distribution is consistent
with half of the sample having flat cores and the remaining half
showing a gradual decline from 0 to $-0.8$ (dlog$\Sigma/$dlog$r$). We
deproject the surface brightness profiles in a non-parametric way to
obtain luminosity density profiles. The distribution of luminosity
density logarithmic slopes show similar features with half of the
sample between $-0.4$ and $-1.8$. These results are in contrast to our
theoretical bias that the central regions of globular clusters are
either isothermal (i.e. flat central profiles) or very steep
(i.e. luminosity density slope $\sim-1.6$) for core-collapse
clusters. With only 50\% of our sample having central profiles
consistent with isothermal cores, King models appear to poorly
represent most globular clusters in their cores.

\end{abstract}

\keywords{globular clusters:general, stellar dynamics}

\section{Introduction}

\subsection{Surface Brightness Profiles}

Globular clusters (GC) are nearby isolated and relaxed systems, which
makes them good laboratories to study stellar dynamical processes. As
a first step for any dynamical model, we require a measure of the
surface brightness profile. Dynamical processes such as core-collapse,
influence of a central black hole, and the physics of the initial
collapse \citep{bah77,coh80,gne99} will influence the central surface
brightness profile, while tidal influences and evaporation leave
noticeable effects at larger radius. The standard view is to assume
that the central regions are isothermal and the outer regions are
tidally truncated by the galaxy. King models \citep{kin66,mey97}
provide a theoretical base for their study. However $\sim$20\% of the
galactic globular clusters show deviations from King models by having
steeper central surface brightness profiles \citep{djo95}. These
clusters have historically been called post core-collapse since this
steeping of the central profile is the expected behavior during
core-collapse \citep{coh80}. Given the large amount of data collected
from the Hubble Space Telescope ($\it{HST}$), our goal here is to
characterize the central profile in a non-parametric way, thereby
testing whether the cores are in fact isothermal or consistent with
the expected post core-collapse morphology.

The surface brightness (SB) profile provides a fairly simple way to
obtain the mass distribution through deprojection; therefore, reliable
SB profiles of any stellar system are necessary for detailed dynamical
modeling. In the case of globular clusters, most dynamical studies use
parameters such as central surface brightness and half light radius
obtained from King model fits to the observed SB. \citet{tra95}
provide the most complete catalog for GC radial profiles. This catalog
contains profiles constructed from ground-based images using a
combination of star counts and integrated light; they also provide
King model fits to determine core radius, concentration and central
surface brightness. It is worth noting that the concentration is the
only parameter they obtain directly from the King model fit; the other
two parameters are obtained from a Chebychev polynomial fit to the
photometric points. They report uncertainties from a variety of
sources, some are relevant to the outer part of the profiles like sky
brightness determination, while others are particularly important for
the inner parts of the profile such as center determination and
crowding correction for star counts. They report a seeing of $2-3$
arcsec for the observations. While this catalog is extremely useful
for analyzing the outer parts of the SB profile, it is necessary to
update the data for the innermost regions using $\it{HST}$'s
resolution. Another study using ground-based images in the U filter is
performed by \citet{lug95} on 15 core-collapse clusters. They fit pure
power-law and modified power-law (which allows the existence of a
core) to the central surface brightness of these objects. They find
that nine have unresolved cores, three have marginally resolved cores,
and three have clearly resolved cores. The average slope of the
power-law fits is $\sim-0.8$. They conclude that clusters in their
sample, with the exception of one object (NGC~6752), have cores\vspace{5pt}

consistent with expectations for a post-collapse bounce.

Some specific clusters have been studied in more detail. Particularly,
M15 has been the subject of many studies trying to obtain a reliable
radial profile (either in light or in star counts) near the
center. \citet{lau91}, using WFPC, claimed to see a core of
1.1\arcsec; later analysis by \citet{gua96} using WFPC2 found a steep
cusp into the smallest resolution element with a slope of~$-0.7$. This
result is similar to that of \citet{sos97a} using FOC images. Our
results agree with those of \citet{gua96}, as discussed in
Section.~4.3.3. Less detailed studies have obtained SB profiles from
$\it{HST}$ images for M30 \citep{yan94,sos97b}, NGC 6397 \citep{kin95}
and NGC 6752 \citep{fer03}.

In this paper we present surface brightness profiles from $\it{HST}$
images for 38 Galactic globular clusters. In section 2 we describe
analysis of our simulated datasets that allow us to optimize the
extraction of the surface brightness profile. In section 3 we describe
the data acquisition, estimation of the surface brightness profiles
and their uncertainties. In section 4 we discuss the results focusing
on the central slope values.

\subsection{Effects of Dynamical Evolution on the Surface Brightness Profile}

Core collapse is thought to be the process responsible for why radial
profiles deviate from King profiles, therefore, we briefly discuss
it. Core collapse occurs when weak gravitational interactions between
stars drive the central density of the cluster to larger values while
the core radius decreases. This process can be separated in two
stages. First, close encounters drive stars to the halo of the cluster
eventually causing them to evaporate, and the core shrinks due to
energy conservation. This process alone drives core-collapse over long
timescales. A second process, the energy exchange between the outer
halo and the inner core, accelerates the timescale for core
collapse. Mass segregation from two-body relaxation drives energy from
the core to the outer halo, and increases the velocity dispersion in
the core while it contracts.

A number of simulations have been carried out to provide a detailed
description of core-collapse using both N-body codes \citep{mak96} and
numerical integrations of the multi mass Fokker-Planck equation
\citep{bre94, mur90, cher90, coh89}. They all show that the projected
density will have a shallower central slope for the lower mass stars
compared with the high mass stars, although the precise slope for the
visible stars depends on initial conditions. When the presence of
binary systems is included in the simulations, it is seen that they
have important effects on core collapse evolution. The presence of
primordial binaries has the effect of delaying core collapse, but even
if there are no primordial binaries present in the cluster core, hard
binaries are formed by three body encounters once core collapse
begins. These binaries act as a energy source for the core, cooling
it. This in turn reverses the contraction process and produces an
expansion. Eventually, the core contracts again and the whole process
happens periodically, which results in what is known as ``gravothermal
oscillations''. At this stage, the cluster successively goes in and
out of core collapse. It is shown in these simulations that
core-collapse occurs on a very short time scale. The core quickly
re-expands and spends a longer time in a state similar to pre-collapse
between the successive contractions, but with a much smaller core with
radius of a few percent the half-mass radius.

\citep{coh89} predict a central slope of the luminosity density due to
core-collapse of about $-1.7$ for turn-off stars, including effects
from the present mass function with remnants. However, this slope
obviously depends on the mass function, the stage of core-collapse,
and the spatial resolution of the measurements. For instance,
\citet{mey88} measure a non-zero slope for 47Tuc whereas we find a
nearly zero with our improved spatial resolution (as was found by
\citet{tra95}). Another case is NGC6397, where \citep{lug95} measure a
steep cusp from data at large radii, however at small radii they find
a core (they infer that we are seeing NGC6397 in an expanded-core post
collapse state). Thus, in this case, where you define the radii over
which the slope is measured is crucial. Our goal in this paper is to
measure the slope at the smallest radii possible in order to make the
most systematic measurement possible to compare with theoretical
results. In this way, we are not subject to the particulars of the
theoretical bias for what you assume the surface brightness profile
should have.

\citet{dul97} compare Fokker-Planck simulations to the observed
surface brightness and velocity dispersion profiles for M15. They
conclude that these profiles are consistent with an intermediate state
between core-collapse and re-expansion. In a state of complete
re-expansion, the cluster would show a $\sim1.2\arcsec$ or larger
core, which should be observable. If we assume all clusters with
unresolved cores ($\sim20\%$ of galactic clusters) are in a similar
state and we take into account the fact that during gravothermal
oscillations the core spends a very short time in the collapse state,
then we are catching a very high number of galactic clusters in the
act of core-collapse.

Another dynamical scenario that has been explored as part of the
evolution of globular clusters is the possible presence of a central
black hole. \citet{sil75} suggested that central X-ray sources in
clusters could be produced by gas fed into a 100-1000$M_{\odot}$ black
hole. \citet{bah77} calculated the effect on the stellar distribution
for a cluster if a black hole is present in its center. They predict
the formation of a cusp near the center with a logarithmic slope
of~$\sim-1.75$ for the most massive stars in the 3-dimensional
density, while the limiting slope for least massive stars
is~$\sim-1.5$. The predicted slope of the surface brightness
distribution is very close to that predicted for core-collapse for the
dominant stellar components in the core, but the variation with mass
is less dramatic than for the core-collapse case. Most observable mass
groups would have a logarithmic SB slope~$\sim-0.7$
\citep{sos97a}. \citet{bau04} perform extensive simulations of star
clusters containing an intermediate mass black hole (IMBH). They find
that the presence of the black hole induces the formation of a cusp
whose 3-dimensional density profile has a~$\sim-1.55$ slope. In
projection the slope of the cusp is much shallower, yet different than
zero. Recently, there have been two claims for the presence of a
medium size black hole at the center of two globular clusters. One is
for M15 \citep{geb00,ger02,ger03} and the other one is for the giant
globular cluster G1 in M31 \citep{geb02}. Although this is still a
very controversial subject \citep{bau03a,bau03b}, it is crucial to be
able to differentiate between the two possible dynamical states
(i.e. core-collapse vs. intermediate mass black hole). Having reliable
SB profiles near the center of GCs will be a key part of future
dynamical modeling.

\subsection{Non-parametric Models}

Here, we concentrate on the differences between parametric and
non-parametric techniques for estimating the surface brightness. The
way in which we characterize light profiles has important
consequences for dynamical analysis. The advantage of using King
models lies in the fact that they provide a smooth profile even for
sparsely sampled data, and that they have an analytical
deprojection. However, the quality of the data is now good enough that
it is not necessary to use a parametric profile. Furthermore, small
differences or biases between the parametric fits and the data will be
greatly amplified during deprojection, causing the luminosity density
to be possibly poorly represented by King models.  Parametric fits
have a side effect of underestimating the confidence intervals for
three dimensional distributions, since the range of possible solutions
is always larger for non-parametric analysis than for a parametric
one. The draw-back of not forcing a functional form to the
distribution is that the data always have some amount of
noise. Deprojection involves a derivative of the surface brightness
profile; therefore, any amount of noise will be greatly amplified
during the deprojection. Thus, non-parametric algorithms require some
degree of smoothing, and the reliability of the result depends on the
technique and the amount of smoothing used. Ultimately, there is a
problem of assessing whether the fluctuations in the data are real or
not. This is particularly important when the focus of the study is the
inner parts of globular clusters. In this work, we use a
non-parametric approach to analyze our data, similar to that used for
galaxies in \citet{geb96}.

\section{Simulations}

There have been a variety of techniques used in the literature to
measure radial profiles for globular clusters, both with star counts
and integrated light. We performed extensive simulations in order to
test the reliability of different methods for obtaining accurate
surface brightness profiles, which we describe below.

The two most complete studies to date base results on star counts and
both correct for completeness. \citet{sos97a} use artificial star
tests in order to obtain a SB profile for M15. They add synthetic
stars over their image of M15 and measure the recovery rate of their
photometry software. A problem in this case is that it is hard to know
the effect of the underlying stellar distribution on the results,
since the true stellar distribution is not known. \citet{gua96}
perform simulations over a blank image, controlling all the input
variables. They compare the photometry of input and output stars one
by one, calculate a completeness factor for the number of stars in a
given annulus and construct the SB profile from star counts using
those correction factors. This is very reliable but it does not test
for degeneracy that could arise from different underlying profiles
yielding the same final result. Since our goal is to provide a general
prescription by studying the full range of profile slopes, our method
should not depend on the type of profile for each cluster (i.e. cusp
vs. core). We perform simulations over blank images, thus having
control over the input parameters such as the stellar profile,
luminosity function and total number of stars. Below we outline each
step. We argue that using integrated light is superior for measuring
an unbiased SB profile (compared to star counts) if the cluster
contains a large enough number of stars.

\subsection{Image Construction}

Our goal is to create images that resemble the PC chip as closely as
possible. The first step toward creating images is to produce an input
list of stars. We start with a luminosity function for M5
\citep{jim98} and a desired surface brightness profile. The effect of
mass segregation is not included in these simulations. From the
functional form of these two profiles, we construct a probability
distribution for a star having a certain magnitude (from the
luminosity function) and radial distribution (from the surface
brightness profile). Stars are generated randomly around a given
center from those probability distributions. By performing star counts
in magnitude bins we confirm that our resulting star list represents
the supplied luminosity function. The same test is performed in radial
bins for surface brightness. With this method, we create various
master lists of stars of a given surface brightness profile. Results
with fainter and brighter versions of the luminosity function are
discussed below. We use five different power-law profiles
$\Sigma(r)=r^{-\beta}$, with $\beta$ of 0.1, 0.3, 0.5, 0.7 and 1.0 as
the supplied functions for the surface brightness. We also create
images for a King profile with a core radii of 90 pixels. The images
have 200,000, 50,000, 10,000, and 1,500 input stars within a 200 pixel
radius (20\arcsec\ for WFPC2 pixel scale). Five individual realizations
are created for each pair of input number of stars and profile shape
with the goal of performing statistical analysis.

The images are created using the DAOPHOT \citep{ste87} routine
``ADDSTAR". For a base image, we use an actual WFPC2 image containing
very few stars that are cleanly subtracted. This process results in a
realistic background including cosmic rays and bad pixels. The routine
adds Poisson noise and read out noise as well. The supplied point
spread function (PSF) is constructed from the base $\it{HST}$
image. We do not include spatial variation of the PSF since its
relevance varies a lot for each real dataset. The PSF radius defined
for DAOPHOT when building these images is 9 pixels.

\subsection{Center Determination}

Having a good estimate of the center position of a cluster is crucial
to obtain an accurate surface brightness profile. Using the wrong
center typically produces a shallower inner profile. We design a
technique to measure the center that assumes the cluster is
symmetric. A guess center and a radius from that center are chosen.
The resulting circle is divided in eight segments where we count stars
and then we calculate the standard deviation of the eight number
counts. This same calculation is performed for various center
coordinates distributed around the initial guess center with the same
defined radius. The grid of the centers consists of every five pixels
near the center in all directions and every ten pixels further away
from it. This produces a map of coordinates with a standard deviation
value associated to them. We fit a surface to this map using a
two-dimensional spline smoothing technique developed by \citet{wab80}
and \citet{bat86}. The minimum point in the surface is our chosen
center. The method can be used iteratively until the minimum lies in
the finely spaced part of the grid.

All the simulated images have the center in the same position right in
the middle of the chip. The size of the circle we use for our octants
method is 170 pixels, which is slightly smaller than the radial extent
of the simulated clusters. We calculate the distance between the
measured and the real center (in $x$ and $y$ positions separately) for
each of the five individual images in a given setup, then calculate
the average and the standard deviation. Figure~1 shows the accuracy of
the center measurements for all input power-law and King
profiles. Results are shown for both the x and y coordinates in the
case where we input 10,000 stars. We observe that the largest
deviation is equivalent to 5 pixels for this group of simulations
(0.5\arcsec\ on the pixel scale of WFPC2). The center estimation
improves with the degree of concentration of the cluster and with
increasing number of stars. Similarly, the quality of the estimation
decreases with decreasing number of stars and degree of
concentration. As expected, this method works best when the SB
distribution is not flat in the entire image.

\begin{figure}[t]
\centerline{\psfig{file=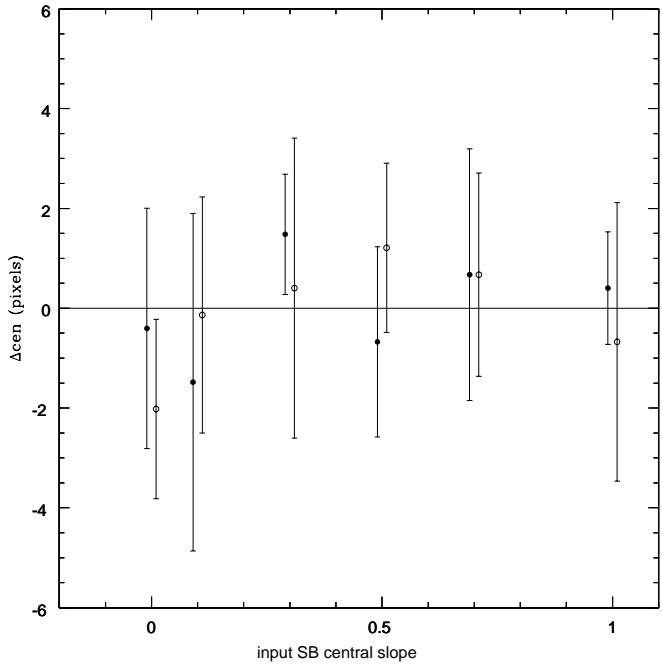,width=9.5cm,angle=0}}
\figcaption{Comparison between the measured and input center for
  various sets of simulations with 10000 input stars. Zero slope is
  for the King profile case and the rest are for power laws. The
  average distance between the actual and measured center is shown for
  the x (solid points) and y (open points) coordinates in pixels. A
  small horizontal offset is introduced for clarity. Error bars are the
  standard deviation of the five individual measurements for each
  case. Each WFPC2 pixel is 0.1\arcsec.}
\end{figure}

\begin{figure*}[t]
\centerline{\psfig{file=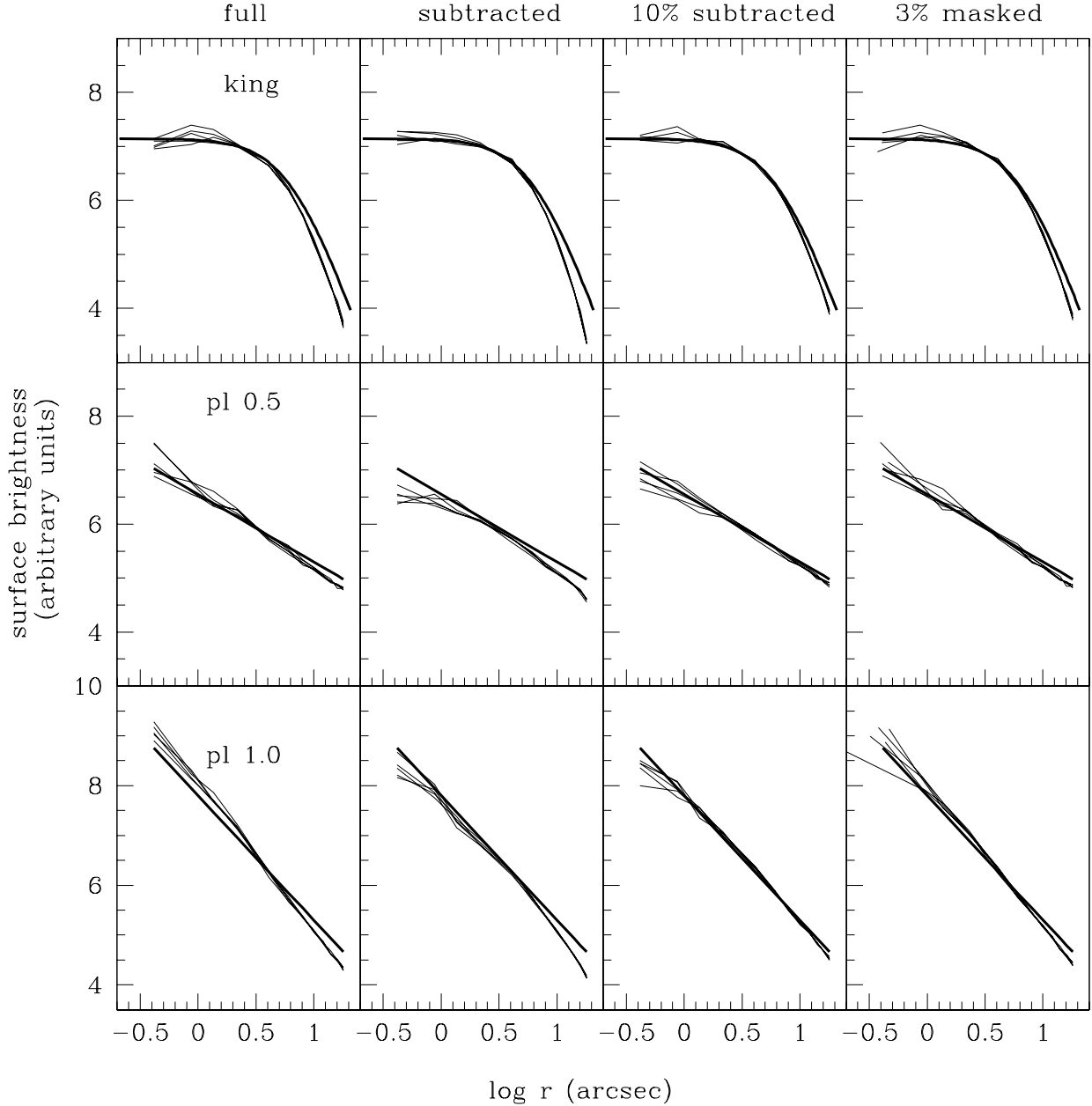,width=17.5cm,angle=0}}
  \figcaption{Surface brightness profiles for three groups of simulations
  with 50,000 input stars. For each case (King profile, 0.5 power-law,
  and 1.0 power-law) five individual measurements (thin lines) are
  plotted against the input profile (thick solid line). The profiles
  are measured from four different images: full, subtracted, 10\%
  brightest stars subtracted and 3\% brightest stars masked. The
  vertical axis is on an arbitrary magnitude scale.}
\end{figure*}

\begin{figure*}[t]
\centerline{\psfig{file=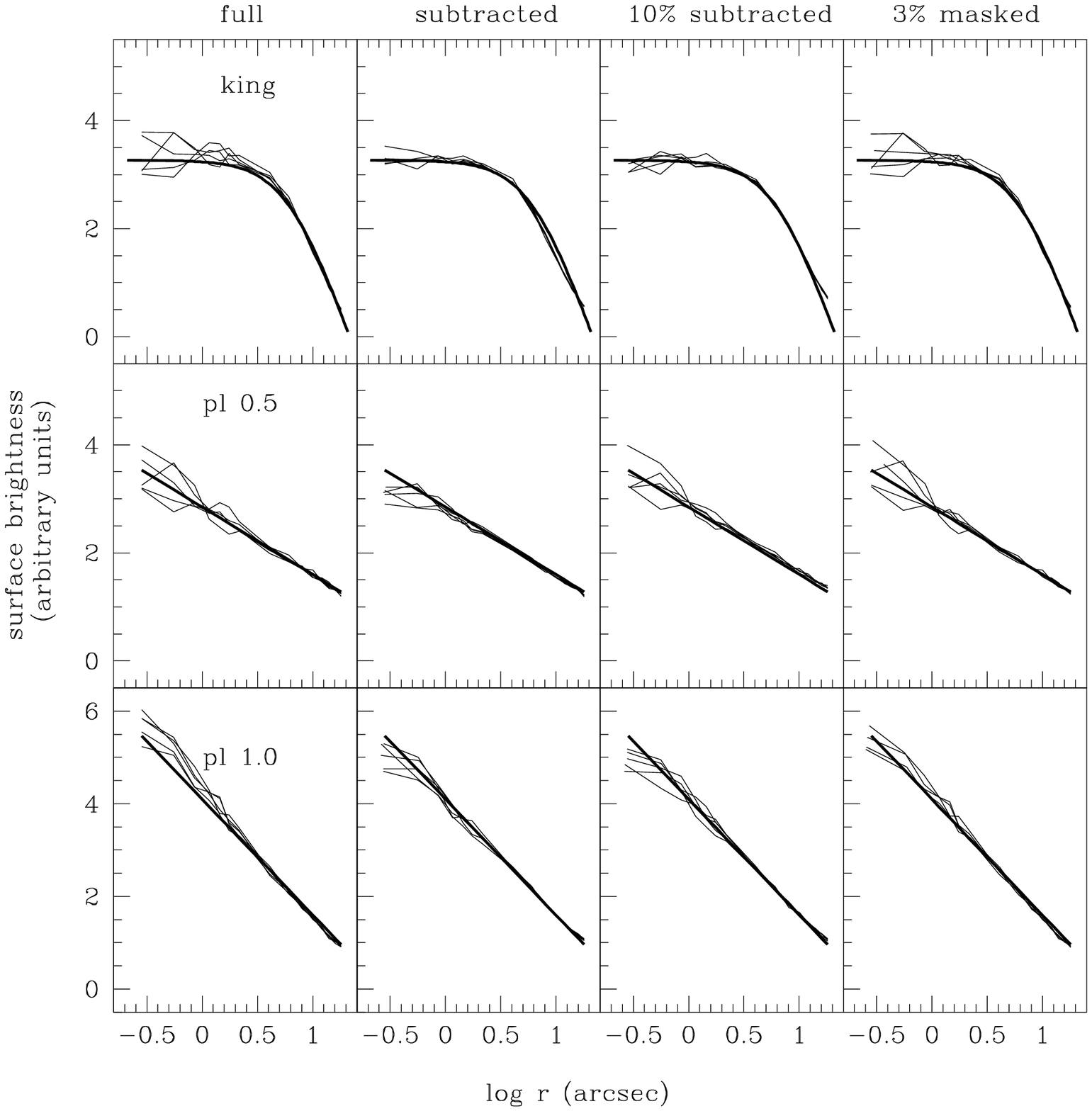,width=17.5cm,angle=0}}
  \caption{Same as previous figure but for simulations of fainter
  clusters. The groups of simulations were constructed by decreasing
  the brightness of stars by two magnitudes.}
\end{figure*}

\subsection{Surface Brightness Profile}

We test several different ways to obtain radial profiles on the
simulated images. The profiles are obtained by measuring both
integrated light and star counts. We note that both techniques have
their advantages and disadvantages; for example, star counts can
measure different radial profiles due to mass segregation while
integrated light cannot. However, we argue that star counts are
significantly less reliable compared to integrated light when trying
to measure the global radial profile. When measuring integrated light,
we use two different statistical estimators---the average and the
biweight \citep{bee90}---to get counts per pixel in a given
annulus. Although the average is an optimally efficient estimator for
central location when dealing with Gaussian distributions, it can be
very biased when the underlying distribution is not Gaussian
(i.e. having outliers). The biweight provides a robust estimate of the
central location (i.e., mean) even when including a significant number
of outliers. Since our images are made from discrete sources, there is
a large number of `background' pixels and a large number of `star'
pixels in each annulus, so the distribution is certainly not
Gaussian. It is important to explore the effect of using a robust
estimator versus using the average.

To measure star counts, we have to first measure the locations and
brightness of all stars using DAOPHOT. We perform PSF fitting star
subtraction on the images using the ALLSTAR routine with the same PSF
we used to construct the image. This does not make the subtraction
perfect since we introduced Poisson noise when constructing the
images. Crowding and read noise have an important effect on DAOPHOT's
abilities to find stars. We observe this by comparing the number of
input versus found stars in each simulated frame. For the 200,000
input stars case, $\sim3800$ are found, while for the 50,000 input
stars case, $\sim3000$. To avoid confusion in the following we refer
to the groups of simulated images by number of input stars instead of
number of detected stars. We now can measure integrated light in two
different images, the original one with all the stars included (from
now on called `full image') and the one with stars subtracted, which
is smoother (from now on called `subtracted image'). We use two sets
of annuli to measure SB from integrated light; one has steps of three
pixels from 1 to 20 pixels radius and the other has steps of twenty
pixels between 10 and 200 pixels radius. The size of these annuli is a
compromise between measuring at the smallest possible resolution and
providing a smooth curve. The radius associated to each annulus is the
midpoint between the outer and inner radii, while the surface
brightness value is the number of counts per pixel divided by the
number of pixels in a given annulus. We find that when using the
average estimator, the profile obtained for integrated light is
slightly biased and is very noisy, while the biweight estimator yields
much smoother profiles with very little bias. The measurements on the
subtracted image always yields a smoother profile than that obtained
from the full one. Figures~2 and 3 show the input profiles together
with the five individual measured profiles for various simulations
with 50,000 input stars. We show a King profile, and the 0.5 and 1.0
power laws. In the cases of concentrated profiles and large number of
input stars, both estimators produce shallower profiles toward the
center for the subtracted images. The reason for this bias appears to
be an over-subtraction near the center of the cluster stars where the
crowding problems are worse. The program subtracts part of the
background starlight as part of the stars which in turn produces a
flatter looking profile near the center of the cluster. We also
observe that the profiles obtained from the full image tend to look
steeper than the input profile for the steepest power laws (inner
slopes in the range $0.5-1.0$) as it can be seen in the leftmost panel
of Figures~2 and 3. This is likely due to the contribution of the
brightest stars near the center where integrated light is being
divided into very few pixels, so the proportional contribution from
the presence of a bright star is much larger near the center than in
the outskirts. This effect can potentially be even larger for real
clusters since they are known to have a degree of mass segregation
\citep{how00}, and therefore they have a larger relative number of
bright stars near the center.

Since our goal is to obtain an unbiased smooth profile, we attempt
alternative ways to measure integrated light profiles. One is to
subtract only a percentage of the found stars, just enough to remove
noise, but not so many that we get over-subtraction problems. We test
for different percentages and compare them with a histogram of found
stars in order to assess which stars are contributing to the observed
bias. After extensive testing, we conclude that subtracting $\sim$10\%
of the brightest stars is optimal. This normally subtracts the giant
and horizontal branch stars leaving most of the main sequence. Another
approach consists of masking a smaller percentage of bright stars. We
choose a masking radius of 5 pixels; this takes care of a large
portion of the light in each star, but it is small enough to avoid
having too few pixels to sample in the central regions. In this case
we obtain profiles with some amount of noise, but we eliminate the
over subtraction problem. By eye inspection of the profiles (Figs.~2
and 3), it appears that the subtracted or the partially subtracted
profiles are the least biased and/or least noisy way to recover the
input profile for the shallower power-laws, while the masked profile
is optimal to recover the higher power-laws.

\begin{figure*}[t]
\centerline{\psfig{file=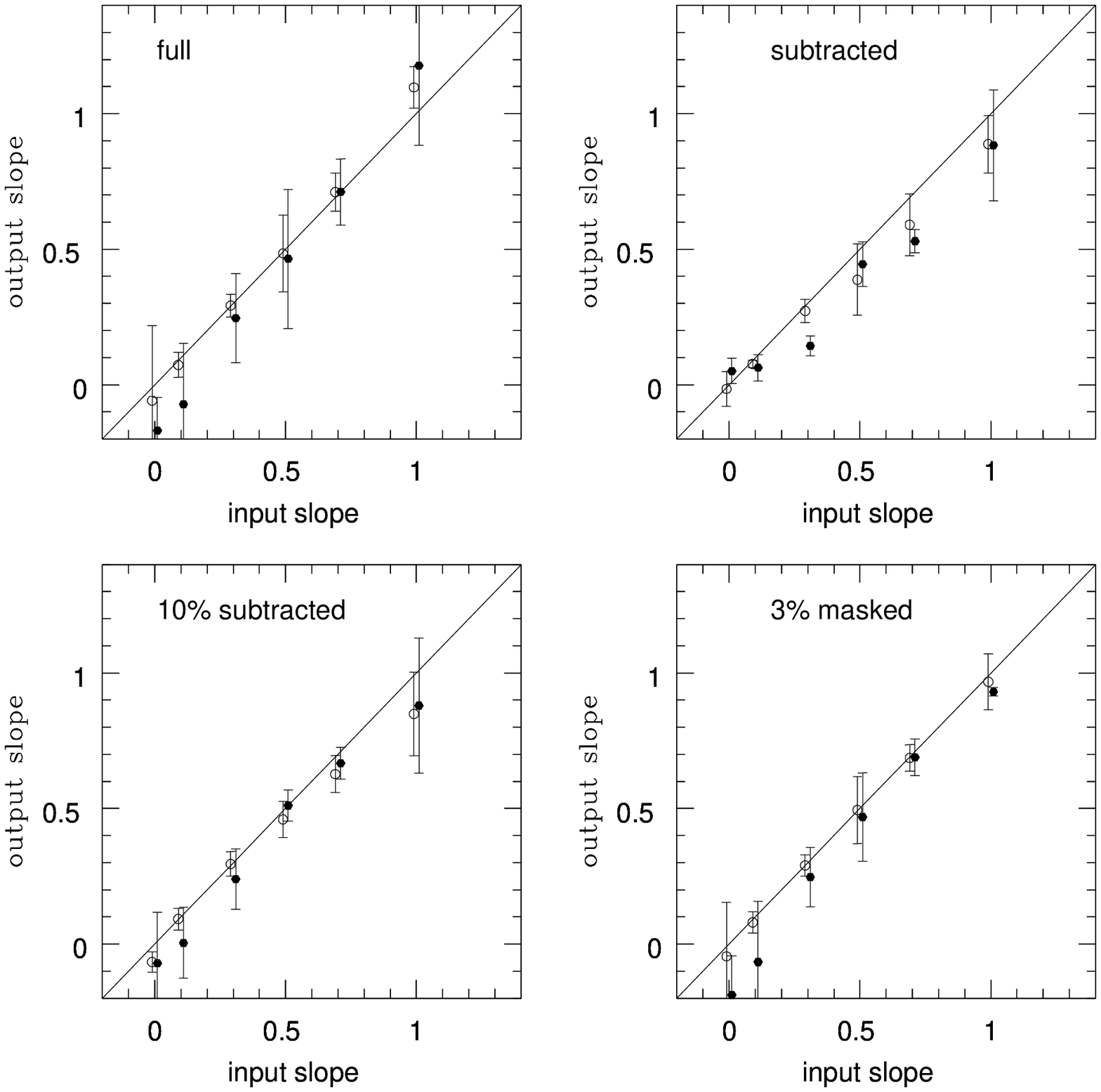,width=18cm,angle=0}}
  \caption{Input versus measured surface brightness slope for two
  groups of simulations. The open points show results for 10,000 input
  stars, the solid points show those for 50,000 input stars. A small
  horizontal offset is introduced for clarity. We show the average
  measured slope of the five individual profiles for each case. Error
  bars represent one standard deviation for the five measurements.}
\end{figure*}

We test the effect of changing the faint end of the luminosity
function for the steepest power-law case by decreasing the number of
faint stars. Our goal is to explore the effect of a change in
background light on the central part of the profile. We normally use a
luminosity function that rises all the way to stars 6 magnitudes
fainter than turnoff stars ($\sim 18$ mag). We change this to a flat
distribution for the faint end (21-24 magnitudes), therefore having a
lower contribution from background light. We find that the effect is
negligible on the final profile; the central shape of the measured SB
profiles was not affected by this change. Therefore, we conclude that
the background light from very faint stars is not an important
contributor to the central SB profile when measuring integrated
light. This result implies that the possible effects of mass
segregation are reduced when we measure the profile from integrated
light, since the contribution to the profile comes from stars with
very similar masses. Therefore, the variations in the radial profile
between the masses of those stars contributing to the integrated light
are minimal. We also test for the effect of distance by using the same
input lists for all cases, but making the stars two magnitudes
brighter in one case and two magnitudes fainter in another. We obtain
smoother profiles for the brighter case and a noticeable bias at large
radii, where the profile is slightly underestimated. For the fainter
case the profiles are noisier, but the bias at large radius seems to
disappear (Fig~3). The over-subtraction related to crowding is
amplified for the brighter case and smaller for the fainter case. M5
(the source of our luminosity function) has one of the brightest
apparent magnitude horizontal branches in the galactic globular
cluster system, so most of the actual observations will be better
represented by the simulated images created with the original and
fainter star lists.

Star counts profiles are obtained in the same sets of annuli we use
for integrated light. Due to crowding near the center of the images,
only a fraction of the faintest stars are detected there. If we
include those stars in the star counts, they tend to flatten the
overall profile, particularly for the steep profiles and large number
of input stars. As a consequence, we decide to use only the 50\%
brightest stars to construct this profile, since this is the limit
where the shape of the input profile is recovered. In general, the
star counts profile, as we construct it, can be used only as a
comparison tool since it is too noisy to provide a robust result. It
is worth clarifying that we do not apply any correction to star counts
due to crowding, which is the normal procedure used by other authors
to obtain star counts profiles in these type of fields. At large
radii, star counts are probably the only way to obtain a surface
density profile. They are certainly the only way to measure the
variation in profiles between different stellar groups within a
cluster, which is something that cannot be measured with integrated
light. However, at small radii, crowding effects severely limit the
usefulness of star counts since they require a significant
correction.

Surface brightness profiles obtained from integrated light can be
noisy for some cases (least concentrated objects, lower
signal-to-noise). Therefore, in order to measure inner slopes, we have
to apply some kind of smoothing and check whether that smoothing
biases our measurements. The smoothing technique is the
one-dimensional version of spline smoothing mentioned in
Section~2.2. It is based on the work by \citet{wah90} and described in
detail in \citet{geb96}. We choose to apply a fixed amount of
smoothing to every profile obtained in order to be consistent. The
central slope is calculated by taking the derivative of the smooth
profile on the few innermost points, which is equivalent to measuring
at a radius of 3 pixels (0.3\arcsec\ with the WFPC2 scale). Results are
shown in Figure~4 for the 50,000 and the 10,000 input star cases. We
plot input versus the average slope measured for the five
realizations. The error bars represent the standard deviation of these
measurements. Results confirm what the eye inspection of the profiles
suggest. The subtracted and partially subtracted images yield a more
reliable inner slope measurement than the full and masked image for
the shallow slope cases ($\beta<0.3$), which seem to have larger error
bars, particularly for the 50,000 input stars case. We confirm that
using the masked profile for those with steep slopes is more reliable;
subtracted and partially subtracted cases tend to underestimate the
slope. In order to further estimate the scatter, we created twenty
images using 50,000 input stars and 0.7 power law. These twenty cases
do not include the five cases already analyzed. The standard deviation
of the slope is slightly smaller for the twenty cases as for the five
images, so the error bars calculated for the five simulations case are
an upper limit.

Besides measuring the core radius (radius where the luminosity drops
by half the central value) we are interested in measuring the turnover
radius (radius of maximum curvature) of the profiles. We do this by
finding the minimum of the second derivative for the smooth
profile. We created groups of simulations with small flat cores to
test if we could detect such turnovers. Our results show that we can
detect cores as small as 1\arcsec\ with our spatial sampling.

Given these results, we use the same four images (full, subtracted,
partially subtracted, and masked) for the real data. If the four
profiles obtained from these images are consistent we take the
smoothest version (in general, this is the masked case). If they
differ near the center (as it is expected for concentrated cases) then
we take the profile produced from the masked image since that is the
one that traces the cusps best. As a general rule we do not use the
profiles obtained from the full image and from star counts because
they appear to be biased for some cases and generally noisier than the
rest.

\subsection{Uncertainties in the Simulations}

The uncertainties in the surface brightness are due to two sources
when using integrated light: the photon noise and the shot noise from
having a limited number of stars (i.e., surface brightness
fluctuations). Thus, in order to get the real uncertainties, we have
to estimate the shot noise from stars. Star counts, on the other hand,
directly recover the appropriate noise, but at the expense of higher
uncertainties due to the difficulties in measuring the individual
stars (i.e., completeness due to crowding).

For the simulations, we have the knowledge of the actual shot noise
since we know the input number of stars. In order to determine how to
include shot noise, we run simulations with the same input parameters
but a different star list. The scatter from these different
realizations then provides the actual uncertainties including both
photon noise and shot noise from the stars. However, with real data we
do not have the luxury of running different star lists; therefore we
have to find a way to determine the shot noise directly. We use a
biweight estimate of the scatter and then apply a correction factor.
The biweight scatter is determined from the scatter in the photon
counts in the pixels for a given annulus. We then compare the biweight
scatter with the scatter of the photometric points between the five
different realizations. The ratio of the real scatter to the biweight
scatter is larger for the simulations with smaller numbers of
stars. Thus, we have to correct the biweight scatter by the
appropriate amount. When using the data, we do not necessarily know
the underlying stellar surface density, making it difficult to
determine the appropriate scaling for the biweight scatter. However,
we use an alternative method that relies on assuming a smooth radial
profile. We discuss this method for real data in Section~3.5. Both
methods give the same range in scalings, implying we have an robust
estimate of the true uncertainties.

Alternatively, we could run proper completeness corrections and
determine the corrected star counts. The standard technique would be
to apply this as a function of magnitude and radius in order to
determine the underlying luminosity function. With that in hand, one
can straightforwardly measure the additional uncertainty due to shot
noise alone. However, this will create an additional source of
uncertainty due to the estimate of the completeness corrections
themselves (the correction factors depend on the underlying
distribution of stars which is precisely what is being measured,
thereby causing a possible degeneracy). Another source of uncertainty
is that star counts will always miss the contribution from the
unresolved stars, which is not an issue for integrated light
measurements. Therefore, we rely on the above approach, and the one
outlined in Section~3.5, where we calibrate the uncertainty estimates
for the actual data with the simulations presented here. Since the
simulations demonstrate that we recover the central shape accurately,
our adopted approach is reliable.

\section{Data acquisition and analysis}

\subsection{Sample}

$\it{HST}$ has imaged a large fraction of all globular clusters in our
galaxy. \citet{pio02} obtained color magnitude diagrams for 74
galactic GCs from WFPC2 images. In addition, \citet{mac03a,mac03b}
obtained surface brightness for clusters in the LMC (53 objects) and
SMC (10 objects), which we analyze in a future paper. Based on our
simulations only a subset of the Piotto snapshot observations will
provide a reliable SB profile since a minimum number of counts are
needed in the frame. Given the distribution of concentration, total
magnitude, and apparent magnitude of the horizontal branch only a
fraction of the imaged clusters are useful. The requirement is to have
enough total counts in the frame. This can be achieved by the cluster
being near (bright horizontal branch), containing a large number of
stars, or being very concentrated (but not dominated by one star). In
general, detecting stars six magnitudes fainter than the horizontal
branch with a signal to noise of 20 is a minimum requirement for low
concentration clusters. This criteria can be relaxed for highly
concentrated clusters (c$>$2.0) and those with a large number of stars
($M_V<-7.5$). Using these criteria we gather from the $\it{HST}$
archive a sample of 38 GC imaged with WFPC2. It is ideal to perform
the study with images in U-band (F336) since giant stars contribute
the same amount of light as main sequence stars at this wavelength,
thus minimizing shot noise.  Unfortunately, there are few images
available with enough signal in U-band. Our selection criteria is
using images observed in either V (F555), R (F665), or I (F814)
filters and to have an exposure time of at least 100 seconds, although
most of the images have exposure times over 500 seconds (see Table
1). After testing for consistency between filters (details below), we
realize we can also include images in the U filter with long enough
exposure times ($>1000$ sec). The field of WFPC2 is 2.6\arcmin\ in
size, which is adequate to measure out to $\sim$2.5 half-light radius
of most clusters but not out to the tidal radius. The scale of the CCD
is 0.1\arcsec/pixel for the WF chips an 0.046\arcsec/pixel for the PC
chip.

We use the WFPC2 associations from the Canadian Astronomy Data Center
website\footnote{http://cadcwww.dao.nrc.ca/}. These images are
spatial associations of WFPC2 images from a given target, normally
coming from a single program. The individual raw data is processed
through the standard calibration pipeline, grouped in associations and
combined. The available data is a multi-group image with the images
for the three WF and the PC chips.

\subsection{Image Processing}

We analyze the WFPC2 images using the same method applied to simulated
images described on the previous section.  Once we have an individual
image for each chip, we trim the edges due to increased noise
there. We use the ``FIND" task on DAOPHOT to obtain a list of stars,
followed by the task ``PHOT" to perform preliminary aperture
photometry. We construct a PSF for each of the four chips. After
extensive testing for methods to automatize this process, we conclude
that the best way to obtain a reliable PSF subtraction is to choose
PSF stars by hand. A single bad PSF star has an important effect on
the quality of the subtracted image. Once we have the list of PSF
stars, we use an iterative procedure where a preliminary PSF is
constructed, neighbors to the PSF stars subtracted, and recalculate
the PSF. We also test constructing a PSF with spatial variations but
in the end this does not have an effect on the quality of the measured
profiles, so we construct a constant PSF for all images. In the end we
have an image for each chip with all the stars subtracted and only
background light remaining. We also produce images with only 10\% of
the brightest stars subtracted, and 3\% of the stars masked as
described in Section~2.3. A geometrical transformation of the
individual images produces a mosaic image.  We end up with four mosaic
images for each cluster; one with all stars included, all stars
subtracted, 10\% of the stars subtracted, and with 3\% brightest stars
masked.

\subsection{Cluster Center Determination}

To determine the cluster center, we first transform all found stars to
a combined coordinate list. We use transformations identical to those
applied when making the mosaiced frame. With this master list we
calculate both the center and radial density profile from star
counts. The center is obtained with the method described in Section
2.2. The first guess center is made by visual inspection of the image
when possible, then iterated until we find the best center. For the
least concentrated cases ($\sim30\%$ of the clusters) we have to make
our initial guess using Digital Sky Survey images with a larger
field. The radius for our method is chosen so that all the stars
counted would lie within the chip containing the center of the cluster
and it is always larger than the core radius. For two of the clusters
(NGC~6624 and M69), the center is too close to the edge of one of the
chips, so we had to use stars on the adjacent chip to find the
center. For another case (M13) the core is larger than the chip so we
also had to use stars in the adjacent chips. Three of the clusters
have too big and sparse cores for this method to work (NGC~5897, M10,
NGC~6712). For these cases we used the center indicated in the Harris
catalog \citep{har96}; these cases are marked in Table 1 with an
asterisk. It is worth mentioning that the sky coordinates reported in
our table come directly from the WCS information contained on the
header of the images, so they should be used only in the context of
that specific image. We have noticed that the sky coordinates of a
specific star can change by as much as 1.8\arcsec\ in two images with
different headers due to $\it{HST}$ pointing uncertainties. The
differences between our center coordinates and those contained in
Harris' catalog are discussed in Section 4.1.

\subsection{Surface Brightness Profiles}

With the center from the stellar data, we obtain a surface brightness
profile from the integrated light in each of the four images. We do
this by measuring a biweight (see Section~2.3) of counts per pixel on
a given annulus, and then dividing that over the total number of
pixels on the annulus. We use a different set of annuli for each
object. Our goal is to obtain the best possible spatial resolution,
while keeping the noise as low as possible. For each case there is a
trade off between these two quantities. We also bin in order to have a
good sampling around the `turnover radius'. In the end we define three
sets of concentric annuli: 3-7 pixels steps at 1-20 radius, 6-15
pixels steps at 15-35 radius, and 30-60 pixels steps extending the
radial coverage to 800.

When we calculate the star count profile, as the analysis in
Section~2.3 suggest, we cut the PSF subtracted star list to keep only
the 50\% brightest stars when we construct the profile. Stars are
counted in the same annuli as the integrated light measurements and
divided by the number of pixels in each annulus. In the end we obtain
five profiles for each cluster from the full, subtracted, partially
subtracted, masked images, and star counts. For most clusters the SB
profile obtained from the full image or from the star counts are
noisier compared to the others, so we never use them as the final
profile. For the cases with steep cusps, there is always a difference
near the center between the masked, partially subtracted, and
completely subtracted profile, as observed for the simulations. In
this case we always choose the result from the masked image since
simulations show this is the least biased. For the cases where the
masked, subtracted and partially subtracted profile have the same
shape, we take the masked profile if it has the same amount of noise
as the rest; but for a few cases we take either the subtracted (M3,
NGC~6287, M92, and NGC~6388) or the partially subtracted (47Tuc, M79,
M5, M80, M62, M9, M69, and NGC~6712) because they are smoother. These
are all cases where the central profile is nearly flat.

If a very bright or saturated star lies near the center of the cluster
it can have an important effect on the final profile, either because
the PSF subtraction is poor or because of the presence of diffraction
spikes that are not included in the PSF. From tests where we mask
bright stars near the center of a cluster, we determine that they only
affect the shape of the final profile if they are within 1 arcsecond
from the center. M70 is the only case where we had to mask a bright
star located withing this region. Since this is a saturated star, we
also mask the diffraction spikes. This occurs at the cost of
decreasing spatial resolution because we cannot use the inner 5 pixels
for our measurements.

\begin{figure}[t]
\centerline{\psfig{file=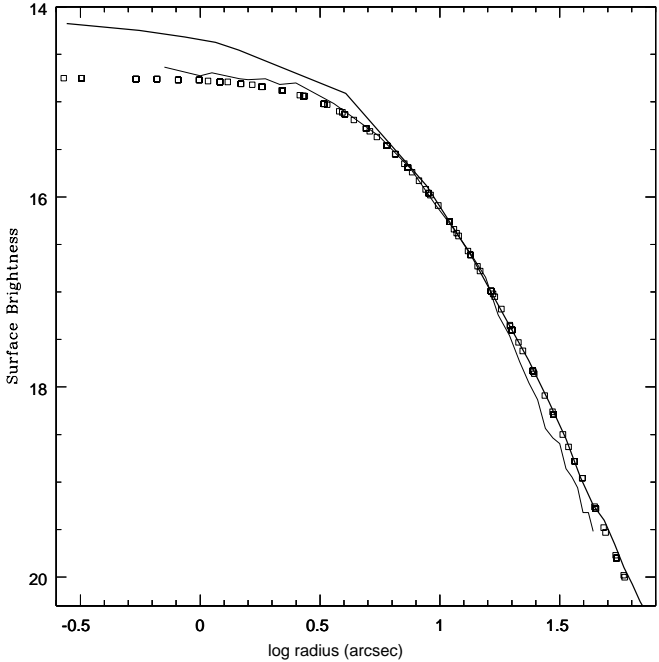,width=9.5cm,angle=0}}
  \figcaption{Surface brightness profiles for M54. The vertical axis
  shows a magnitude scale. The squares show the Chebychev polynomial
  fit from Trager's catalog. The thick line shows the profile obtained
  from a WFPC2 image with our method. The thin line shows the profile
  obtained from the same WFPC2 when it is binned and convolved to
  mimic a ground based image. The change in the central SB profile is
  due primarily to improved spatial resolution from $\it{HST}$.}
\end{figure}

The profiles that we recover sometimes differ greatly from previous
ground based data. In order to check that this is due to improved
spatial resolution, we bin one of our high signal to noise WFPC2
images to the reported pixel scale (0.4\arcsec) of the data used in
Trager's catalog \citep{djo86}; we then convolve this image to account
for the typical seeing reported for the observations in Trager's
catalog ($\sim 2''$). We compare the profile obtained from this
binned-convolved image with that obtained from the $\it{HST}$
image. Fig~5 shows that the effects of pixel scale and seeing can hide
a shallow cusp that can be well measured with $\it{HST}$
resolution. While this effect has been well demonstrated for galaxies
\citep{lau95}, it has not been appreciated for clusters. The profile
obtained from binning and convolving the image lies on top of the
Chebychev polynomial fit to Trager's photometric points, while the
$\it{HST}$ profile is brighter near the center.

Another important test is to check for a possible filter dependence of
the shape of the SB profile.  M80 has observations available on F665
(780~sec), F555 (96~sec) and F336 (11,000~sec) filters.  Figure~6
shows the SB profiles for each. We observe that the three profiles are
consistent throughout the radial range, and they differ by the same
amount from Trager's Chebychev fit.  Thus, we use results from various
filters.  Obviously, color properties will cause some
variations. \citet{gua98} report a $\Delta B-V\sim 0.3$ mag from
1\arcsec\ to 10\arcsec\ for M30. Since our main objective is to obtain
the central slopes, the small color gradients will have little effect.

We also require surface brightness profiles extending out to large
radii. The WFPC2 camera only covers the central region, and we must
rely on ground-based observations. For this, we use the Chebychev
polynomial fit to the photometric points from \citet{tra95}. We use
our photometric points for the inner $\sim20$\arcsec\ and the Chebychev
fit for the outer region. In a few cases the agreement between the
polynomial fits and our results is good throughout, but for many cases
there are discrepancies. We normalize the $\it{HST}$ surface
brightness to the ground based data by matching the two enclosed light
profiles, calculated by integrating the SB profiles. As expected, the
enclosed light curves differ in shape at small radius, but for most
clusters, the curves have the same shape at large radius. Regardless
of which filter is used to construct our profiles, the fact that they
are all matched to photometric points in V and that the profiles are
consistent between filters, brings all our photometric points to V
magnitudes. There are a few clusters for which our normalization
procedure is complicated (M70, NGC~6535, and M15). They all show a
very steep profile through the entire radial range available in our
images; since the ground based data show a core, the shape of the
enclosed light profile obtained from $\it{HST}$ doesn't quite match
that of the ground-based case. Uncertainty in this normalization does
not affect the shape of the inner profile, but it will affect the
value of central surface brightness.

\begin{figure}[t]
\centerline{\psfig{file=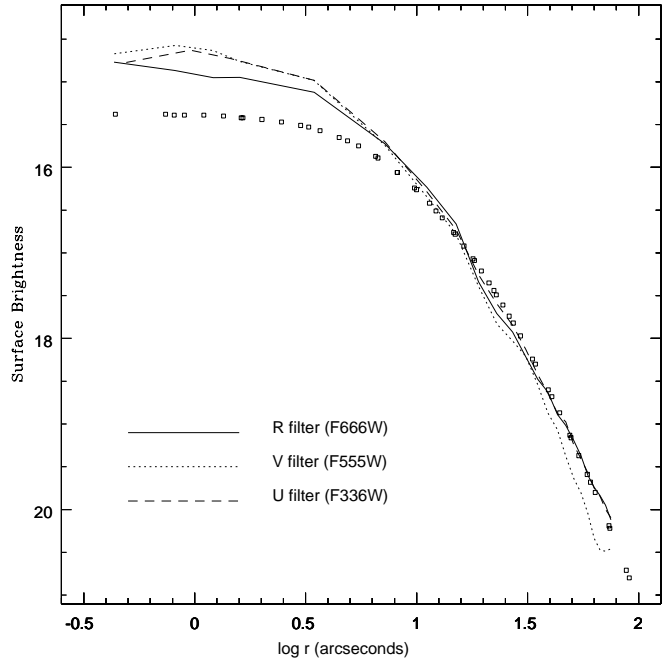,width=9.5cm,angle=0}}
  \figcaption{Surface brightness profiles for M80. The different lines
  show profiles in various filters (F336W, F555W and F665W). The
  vertical axis shows a magnitude scale. The squares show the
  Chebychev polynomial fit from Trager's catalog (measured in V
  band).}
\end{figure}

\begin{figure*}[t]
\centerline{\psfig{file=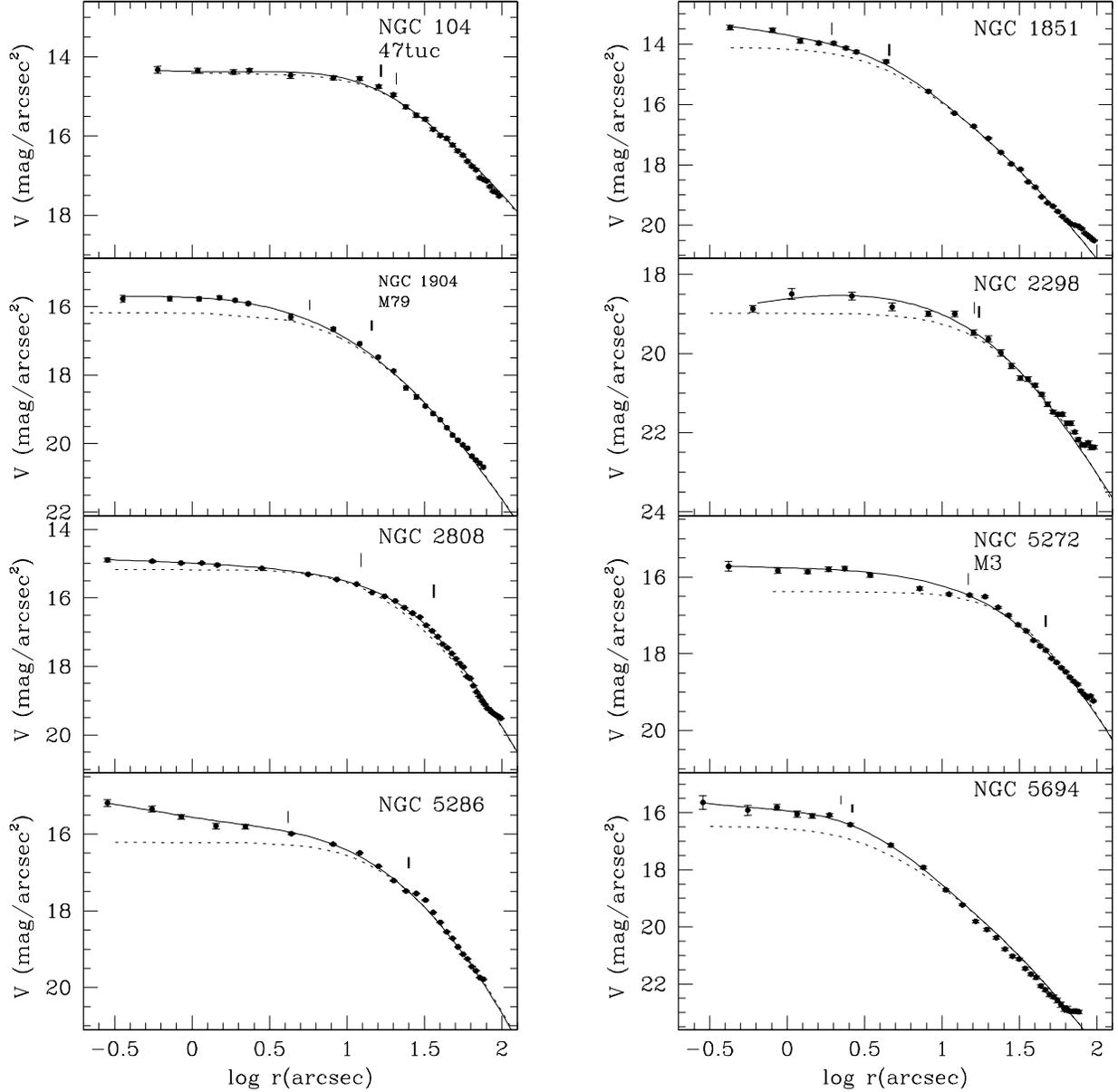,width=18cm,angle=0}}
  \caption{Surface brightness profiles for the entire sample. For each
  cluster we show our photometric measurements (solid points), our
  smooth profile (solid line), and Trager's Chebychev polynomial fit
  (dotted line). The smooth profile comes from a fit to our
  photometric points inside $\sim20$\arcsec\ and the Chebychev fit
  outside that region. For every panel the SB units are V
  mag/arcsec$^2$. We mark the location of the core (thin vertical
  line) and break (thick vertical line) radii. The core radius is
  where the central flux falls by half its value and the break radius
  is where the second derivative of surface brightness with respect to
  radius reaches a minimum.}
\end{figure*}

\begin{figure*}[t]
  \figurenum{7}
\centerline{\psfig{file=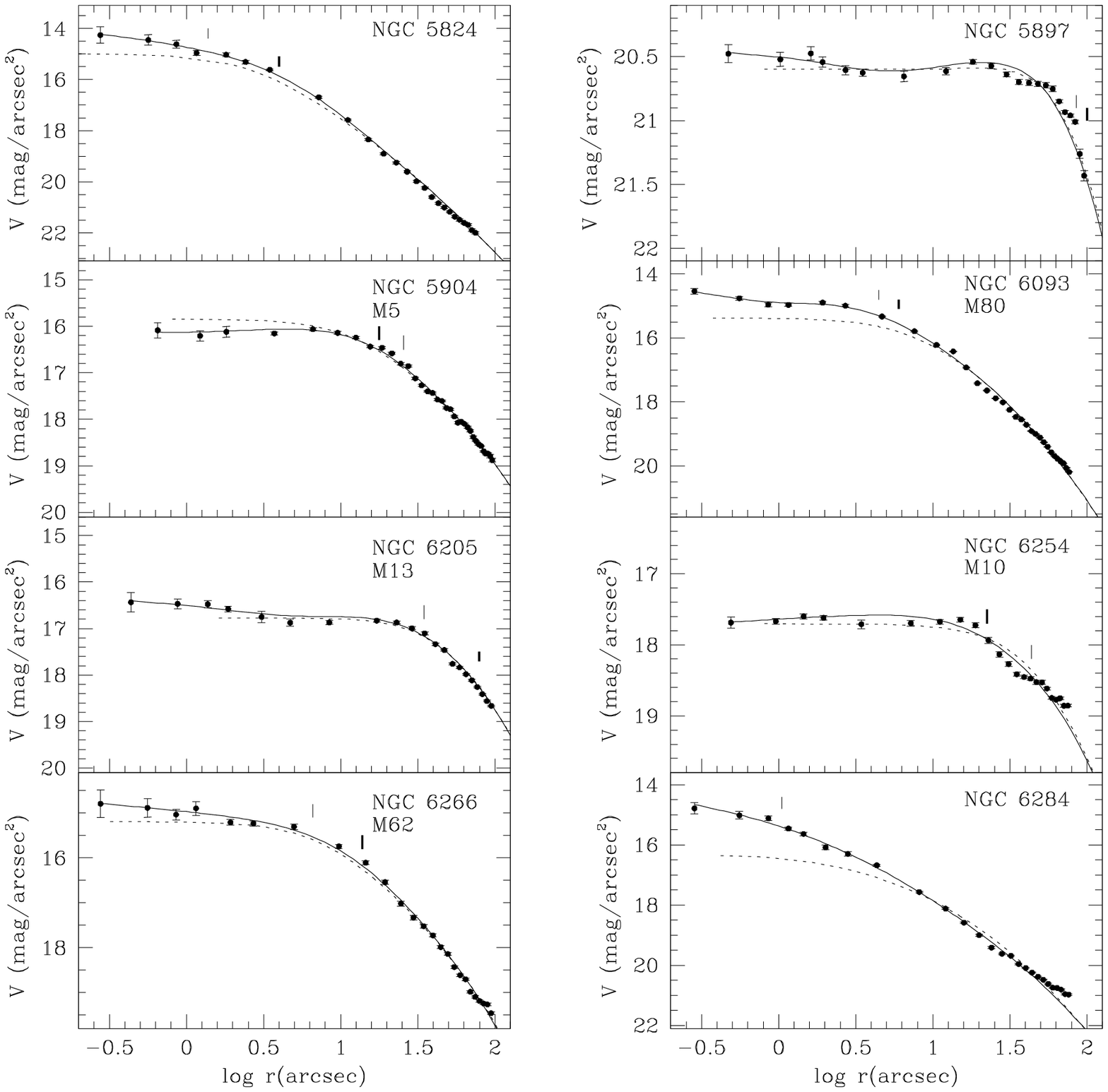,width=18cm,angle=0}}
  \figcaption{continued}
\end{figure*}

\begin{figure*}[t]
  \figurenum{7}
\centerline{\psfig{file=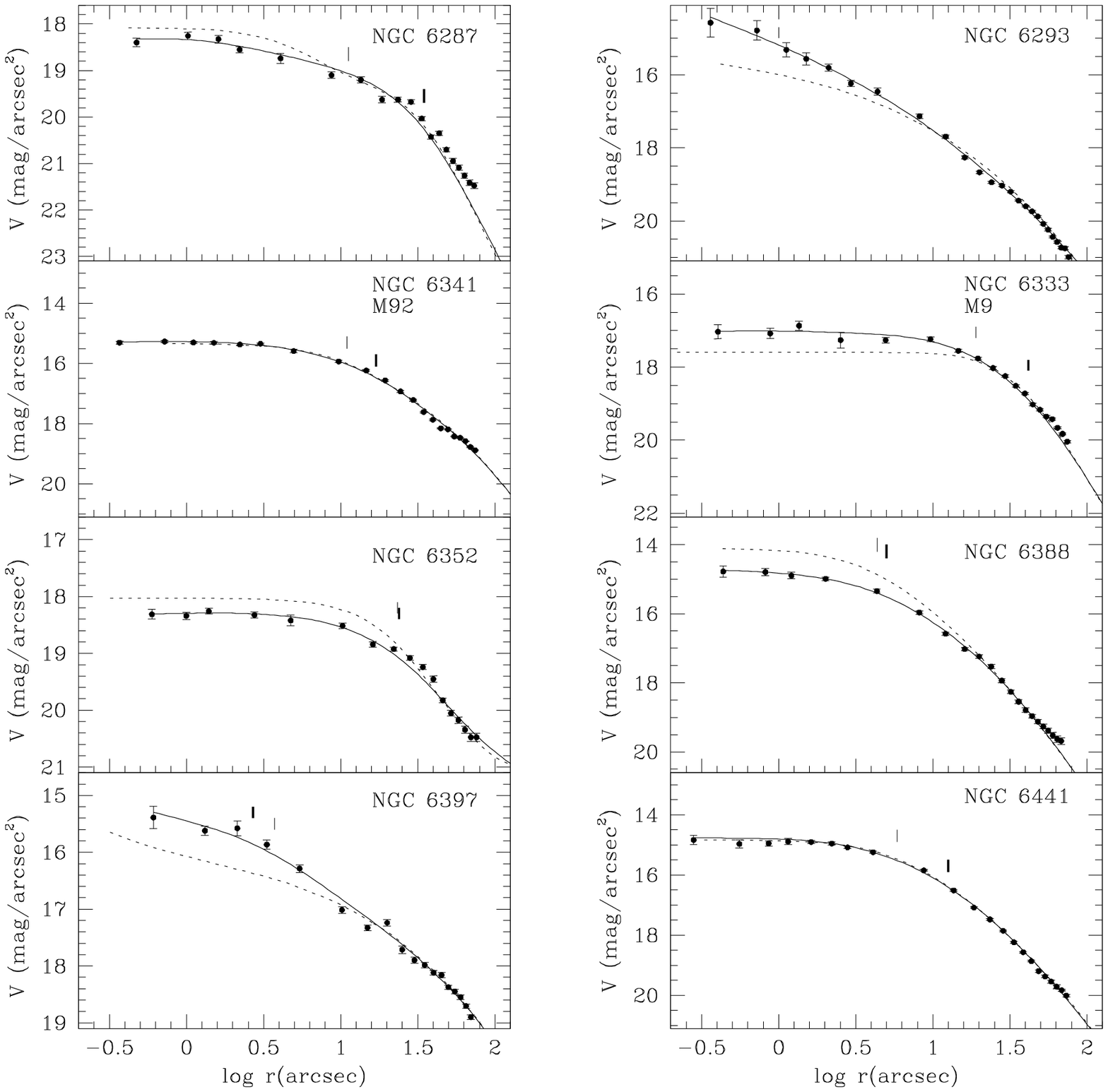,width=18cm,angle=0}}
  \figcaption{continued}
\end{figure*}

\begin{figure*}[t]
  \figurenum{7}
\centerline{\psfig{file=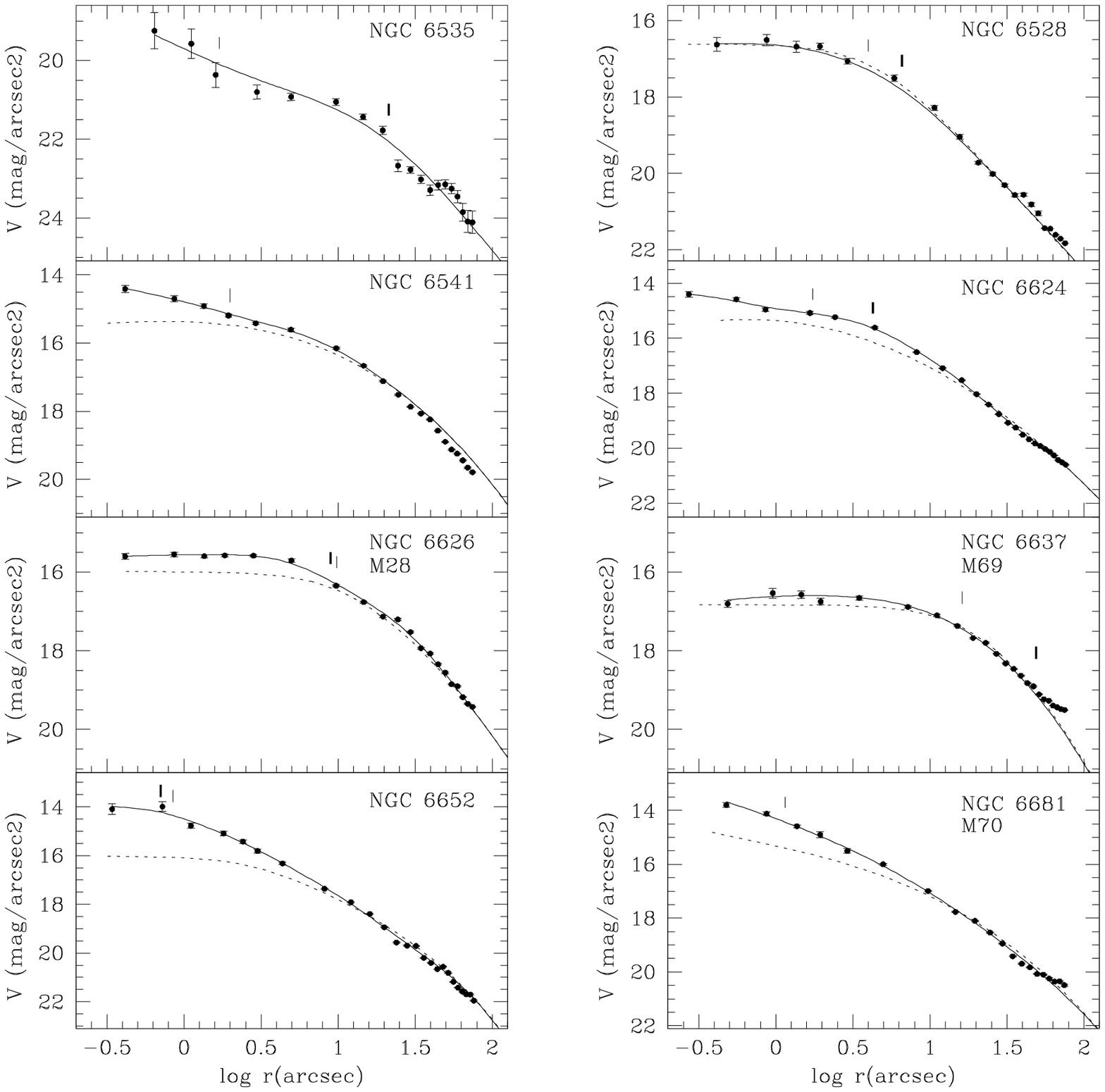,width=18cm,angle=0}}
  \figcaption{continued}
\end{figure*}

\begin{figure*}[t]
  \figurenum{7}
\centerline{\psfig{file=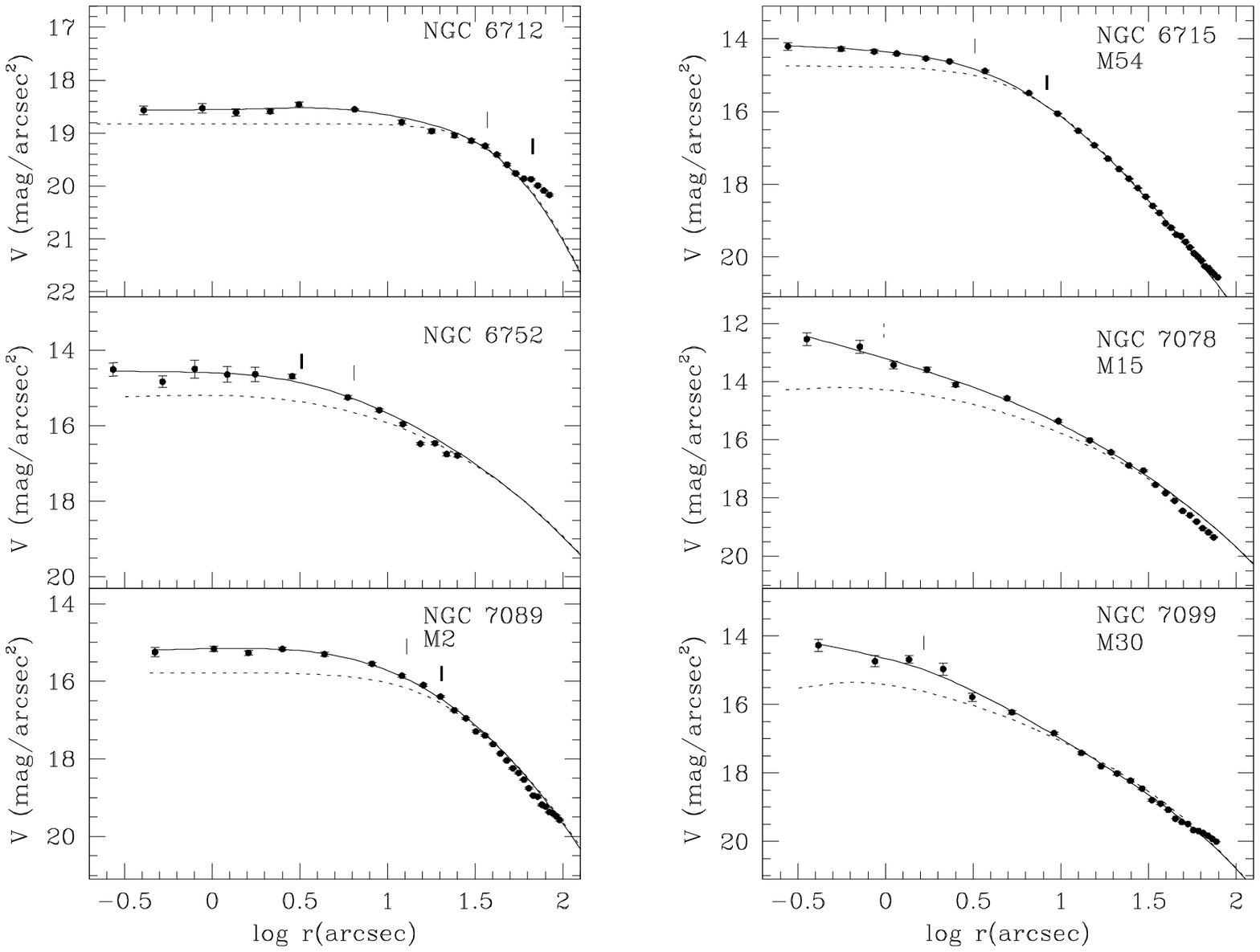,width=18cm,angle=0}}
  \figcaption{continued}
\end{figure*}

\begin{figure*}[t]
\centerline{\psfig{file=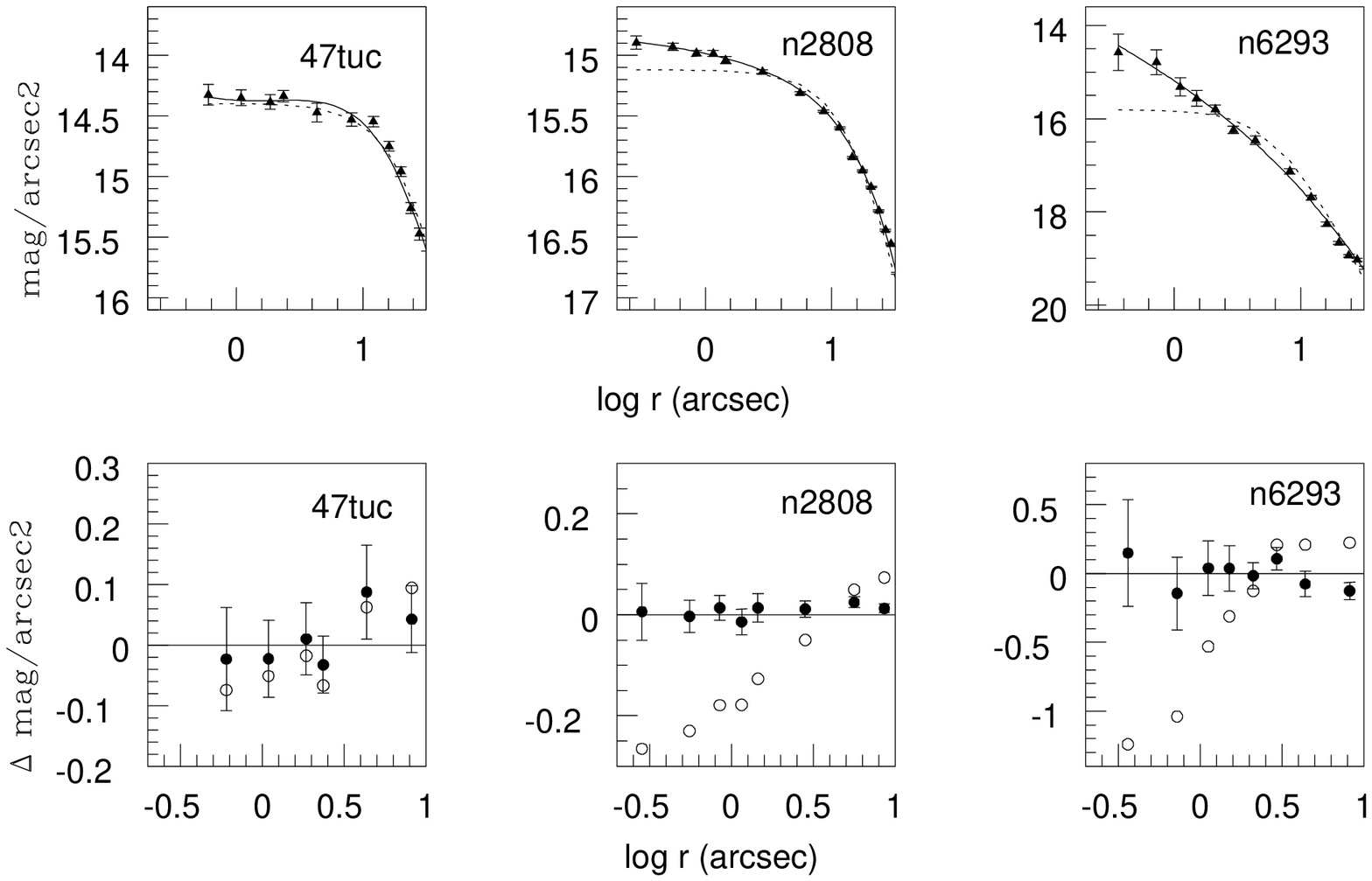,width=18cm,angle=0}}
  \figcaption{Representative single-mass King fits for 47Tuc, NGC~2808,
  and NGC~6293. The top panel shows the photometric points (triangles)
  along with our smooth fit (solid line) and a King fit (dotted
  line). The bottom panel shows the residuals for the smooth fit
  (solid points) and for the King fit (open points).}
\end{figure*}

After normalizing, the final surface brightness profile is a smooth
version of the combination of our photometric points in the center and
Trager's Chebychev fits outside. The smoothing technique is the one
described in section 2.3 for the simulated images. Once we have a
reliable surface brightness profile, we deproject it to obtain the
luminosity density profile. This is done by numerically calculating
the first Abel integral, as in \citet{geb96}. The Abel integral uses
the derivative of the SB profile so any amount of noise in the profile
is greatly amplified. Therefore, we have to apply some amount of
smoothing before deprojecting it (as described in section 2.3). Some
clusters, particularly the ones with shallow cores, yield very noisy
profiles near the center, making the process of deprojection
challenging. For these cases we apply a pre-smoothing process where we
substitute the innermost three or four photometric points by the
average between their two adjacent points. In this way we can apply
the same amount of smoothing to every profile in the sample. For a few
cases, even if we apply the pre-smoothing procedure, we obtain a
surface brightness profile that decreases slightly near the center,
which produces a luminosity density profile with a negative density in
the center and we cannot achieve a proper deprojection. For these
cases, we set the central luminosity density slope to zero (marked
with italics in Table 2).

We measure the central logarithmic slope of the smoothed surface
brightness and luminosity density profiles by taking a first
derivative with respect to the logarithmic radius. In the inner part
of the profile, there is often a range where this derivative is
constant, which implies that the profile has a constant slope in that
region. We take the value of the derivative in this region as the
inner slope for each cluster. The only exceptions are the objects
which have steep cusps; for these cases, the slope changes through the
entire radial range, so we take the value of the innermost points as
the inner slope. Central slope measurements by other authors might be
steeper for a given object, because they tend to fit a power-law in a
more extended radial range (see example in section 4.3.3). For the
cases where the SB logarithmic slope is slightly positive and we
cannot achieve a deprojection, we just assign a zero value for the
slope of the luminosity density. For these cases the values are
written in italics on Table~2. We also measured the values of slopes
in the region outside the core. In this case, since the values of the
first derivative of the profile vary through this radial range we
perform a least square fit to a line for the smooth profile.

Since we are re-deriving SB profiles, we need to measure core radius
as well. Historically, the core radius has been considered as the
radius where the value of the flux falls by half the central
value. The radius often coincides with the radius where the profiles
seem to turn over and change slope, which we call break radius. We
distinguish these two radii for our profiles. The core radius is
calculated by taking the central surface brightness and finding the
radius where the flux falls by half this value. We should note that
the central surface brightness is measured as the value for our
innermost data point, therefore, this value of core radius is
resolution dependent for the non-zero slope cases. We also calculate a
break radius by finding the radius that corresponds to the minimum of
the second derivative of the smooth profile. This is the radius where
the slope of the profile changes by the largest amount, so it can be
seen as the turning point for the curve. Both radii are presented in
Table 2. For the cases with slopes less steep than $-0.5$, where we can
measure a break radius, we compute the ratio of the smallest
resolution radius with break radius. For all cases this ratio is
smaller than 0.15, which means that the break radius is at least 6
times larger than our smallest resolution radius. We plot this ratio
versus the measured value of central SB slope and find no
correlation. In this way we are confident that our reported values for
central slopes in the weak cusps cases are well within the observed
core of the clusters and the slope value is not due to lack of
resolution.

\subsection{Uncertainties for the Data}

In Section 2.4, we describe how we estimate uncertainties for the
simulations, which are based on different realizations where we can
include the shot noise from stars directly. Here we describe the
method we used to calculate the uncertainties for real data and we
calibrate these method against that used for the simulations. We
assume that the underlying stellar radial profile is smooth. Then the
uncertainties of the photometric points should reflect deviations from
a smooth curve in a statistically meaningful way (i.e., have a
Gaussian distribution around the mean value). From the photometric
points, the biweight yields an estimate for the central location and
scale (scatter); this scale value is divided by the square root of the
number of sampled pixels and used as the initial uncertainty for
individual photometric points. We then calculate the root mean square
(RMS) difference between the smooth profile and the data points for
the central region. The ratio of the biweight to the RMS should
represent our lack of inclusion of shot noise from the stars. This
ratio depends on the extent of the radial bins (i.e, the number of
pixels used), therefore we use two different scalings for the
different binning. The average scaling for the inner points is about~2
and about~7 for the outer points. This numbers are consistent with
what we found in the simulations. Thus, we are effectively including
shot noise from stars. The largest scalings occur for sparse clusters
(NGC~6397, NGC~6535 and NGC~6752), as expected.

We calculate the uncertainties on slope measurements from a bootstrap
technique and compare these with the values measured for simulated
images. The bootstrap approach follows that in \citet{geb96}. From the
initial smooth profile, we generate a new profile by generating random
values from a Gaussian distribution with the mean given by the initial
profile and the standard deviation from the photometric
uncertainties. We generate a hundred profiles in this way and measure
the 16-84\% quartiles for the errors. Independently, each cluster is
associated to one of the simulated cases according to its
concentration and number of detected stars, and the standard deviation
from Fig~4 is taken as the uncertainty. These two independent error
measurements agree quite well, which gives us the confidence that the
uncertainties calculated with the bootstrap method are reliable. Table
2 presents this results. The uncertainties for luminosity density
slope measurements is also obtained from the bootstrap calculation. We
do not estimate uncertainties in luminosity density slope for those
cases where we cannot achieve a deprojection. We performed one more
sanity check on our slope uncertainties by measuring the effect of
increasing the uncertainties on photometric points by a factor of
two. From the bootstrap method, we find that the slope uncertainties
increased by a modest factor, less than two, for most clusters. Thus,
the slope uncertainties are not too sensitive to individual
photometric errors.

\section{Results and Discussion}

\subsection{Surface Brightness and Luminosity Density Profiles}

We compare our measured centers (Table 1) with those listed in Harris'
catalog \citep{har96}. For 66\% of the sample the difference is less
than five arcseconds, 24\% of the objects have a difference between
5\arcsec\ and 10\arcsec and only 10\% have a difference larger
than~10\arcsec\ (NGC~1851, M3, NGC~6541 and M2). As mentioned before in
Section~3.2, for three of the clusters (NGC~5897, M10 and NGC~6712) we
used the center listed in the catalog as our center. For the most
concentrated clusters, even a one arcsecond miscalculation of the
center can flatten the central part of the profiles; so this might be
another cause for missing weak cusps in previous measurements.

The SB profiles for the whole sample are shown in Figure~7. For each
cluster we show the SB values measured from the image, the smooth
profile, and the Chebychev polynomial fit obtained by Trager et
al. for comparison. We warn the reader that, as explained in detail on
section 3.2, the photometric points beyond $\sim20$\arcsec\ do not
participate in the fitting of the smooth curve, instead, the Chebychev
fit is used in this region. For most objects the agreement between the
ground based data and ours is very good at large radii
($>10''$). There are a few cases that show disagreement between the
two profiles; these clusters tend to show a steep inner profile
(NGC~6284, NGC~6535, M70, M15), with the largest discrepancies in the
inner 10 arcseconds. As we already discussed (Section 3.2) these
differences may be due to PSF effects. We observe that for 70\% of the
sample the central photometric points are brighter than the polynomial
fit obtained from ground based photometry, sometimes changing the
shape of the previously measured central surface brightness (i.e,
making it steeper). The remaining 30\% agree with previous
measurements or have fainter photometric points near the center. For
the extreme cases, the difference between the central SB value with
previous reports is larger than 1.7 magnitudes (NGC~6284, NGC~6535,
NGC~6652 and M15).

In order to check for any potential biases from our smoothing in the
central regions, we compare with single-mass King profiles
\citep{kin66} fitted to the combination of our photometric points and
Trager's Chebychev fit. For these fits we keep the value of the tidal
radius fixed (from Trager's values) since our data is only in the
central regions. Figure~8 shows representative fits for three
clusters, 47Tuc, NGC~2808, and NGC~6293. For 50\% of the sample, our
smooth profile and the King fit are equally good fits to the data, as
in the case of 47Tuc. For the other 50\%, we obtain either a small
departure from a flat core, as in the case of NGC~2808, or a clear
large departure as in the case of NGC~6293. These departures are
always in the same sense, i.e., the photometric points are brighter
than the King fit towards the center and the deviation increases as
radius decreases. We also performed power-law plus core fits with the
functional form used by \citet{lug95}. We only performed these fits
for the cases that depart from a King profile. The fits are performed
using only the datapoints for the central arcminute, since we do not
expect the outer part of the profiles to be described by a
power-law. For most cases, the power-law plus core fit follows the
same trend as the King fits, but for NGC~6397 and NGC~6652 these fits
are as good as our non-parametric profile. We discuss the details for
each object in Section 4.3.

All of the clusters previously reported as core-collapse show cusps,
with the exception of NGC~6752, which shows a flat core. Only four of
them (NGC~6652, M70 M15, and M30) show a $\sim -1.6$ central
logarithmic slope in luminosity density, which is normally assumed for
objects in this state \citep{bre94}. The rest have slopes between
$-1.2$ and $-1.4$. We consider all objects with luminosity density
slopes more negative than -1.0 to have `steep cusps'; they constitute
34\% of the sample. 24\% show weaker cusps with luminosity density
slopes between $-0.2$ and $-1.0$; many of these have been previously
reported as flat cores or were marked as possibly core collapse on
Trager's catalog. We consider these objects to have 'weak
cusps'. Finally 42\% of the objects in the sample show flat cores
consistent with an isothermal distribution, even when their inner
photometric points are brighter then previous measurements. If we
group the weak cusps with the steep cusps, in total 58\% of the sample
do not show isothermal cores. The presence of so many non isothermal
cores will have important consequences for the dynamical evolution of
the clusters. No dynamical model or simulation predicts this
distribution of slopes for GCs. \citet{gri95} make a detailed study of
large radial structure for 12 galactic clusters. They obtain surface
density profiles from star counts and find that most of the clusters
depart from the King models previously fit to them because they
contain stars in the extra tidal region. This result put together with
the fact that more than half of the objects in our sample are not
represented by isothermal cores leads us to think that King models do
not describe well the surface density profile of many globular
clusters.

Our measured errors for surface brightness slopes are on average 0.1
and the largest is 0.18. For the luminosity density slope the average
is 0.28 and the largest error is 0.54. For the cases with steep cusps,
the error is always under 0.35. Those with measured SB slopes under
$-0.2$ are all $2\sigma+$ detections, implying that they show a
deviation from an isothermal core. Assessing the uncertainties for the
flat cases is particularly relevant since we want to evaluate the
possibility of having positives slopes. Luminosity densities with a
central minimum have been observed in a handful of galaxies
\citep{lau02}. These have been interpreted as two possible scenarios:
one where a stellar torus is superposed on a normal core due to a
recent merger (this is quite unlikely in a globular cluster), and the
other scenario where stars are depleted from the center due to a
binary black hole interaction.  Unfortunately, the uncertainty in our
measurements for cores with positive slope is large enough to include
zero slope.

For each profile on Fig~7, we mark both the core and the break
radius. Seven of the steep cusp cases do not have a measured break
radius because they do not show a clear turning point in the
profile. We observe that for the rest of the sample these two radii do
not always coincide. For all but six cases, the break radius is larger
than the core radius, while for five cases the two are the same. The
core radius that we report is a non-parametric fit as opposed to its
historical value as one of the parameters for King fits.

We also check whether our limited spatial resolution (about
0.3\arcsec) has an effect on being able to resolve a core. We plot the
ratio of our smallest resolution over the measured break radius
against various properties; this ratio is always smaller than 0.2
implying we have at least five resolution elements inside the break
radius for those clusters that have a turn-over in the light
profile. We find no correlations; if all clusters have King-type
profiles with small core radii, we would expect to see correlations.

\begin{figure}[t]
\centerline{\psfig{file=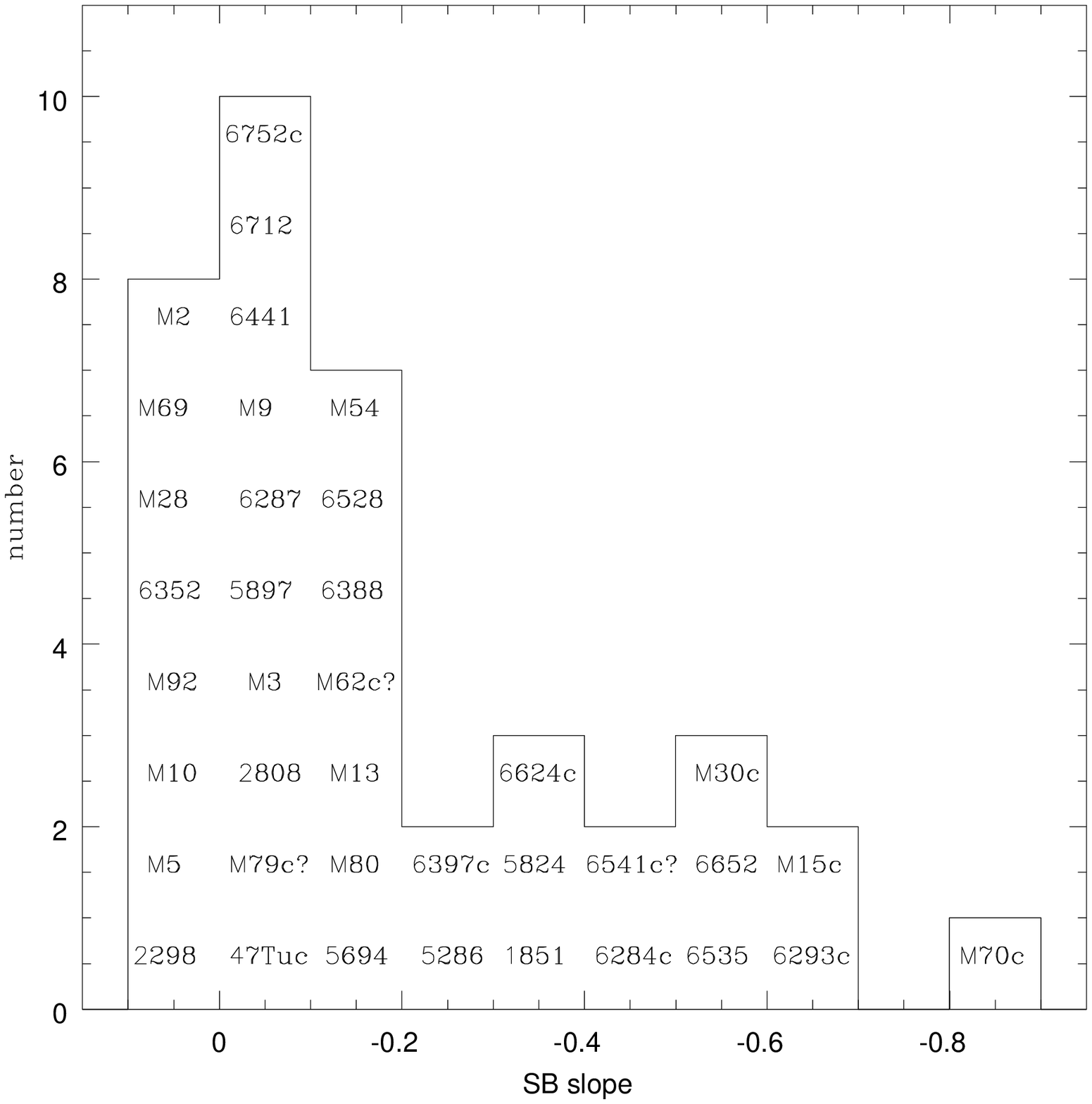,width=10cm,angle=0}}
  \figcaption{Histogram for surface brightness central logarithmic
 slopes. Individual clusters are shown in each bin. The name of the
  cluster is coded according to previously reported dynamical state in
  Trager's catalog. Marked with a `c' for core collapse , `c?' for
  possible core collapse and just the name for flat cores.}
\end{figure}

\subsection{Slopes Distribution and Correlations}

Figures~9 and 10 show histograms of the surface brightness and
luminosity density logarithmic slopes. There is no clear separating
line for two classes of objects, so the sample cannot be cleanly
divided into isothermal and core collapse profiles. Since our sample
is only~$\sim30$\% of the full galactic globular cluster system, we
have to determine potential biases.  Trager et al. classify 16\% of
their sample as core collapse clusters and 6\% as possible core
collapse ('c' with a question mark in his catalog). Our subsample has
21\% objects considered core collapse and 8\% possible core collapse
from Trager et al. Thus, our sample resembles the distribution for the
full sample with a slightly larger number of core-collapse cases. All
but one (NGC~6752) of the objects marked as core collapse fall in our
`steep cusp' category, while those clusters marked as possible
core-collapse are found in all three categories. We find 17 objects
previously classified as flat cores (i.e. classic King models) that
are consistent with an isothermal profile. We can determine the
fraction of clusters that have isothermal cores by comparing our SB
histogram with that expected given our measurement uncertainties for
the clusters that have nearly flat cores. Our average slope
uncertainty is about 0.1. A Gaussian that contains 50\% of the sample
with mean 0 and sigma 0.06 (the average slope error for flat cores)
matches the flat end of the slope distribution very well. The
remaining population ($\sim50\%$ of the objects in the sample) shows a
fairly uniform number of objects between slopes $-0.2$ and $-0.8$. Thus,
only half the objects in our sample are consistent with a King-type
profile.

\begin{figure}[t]
\centerline{\psfig{file=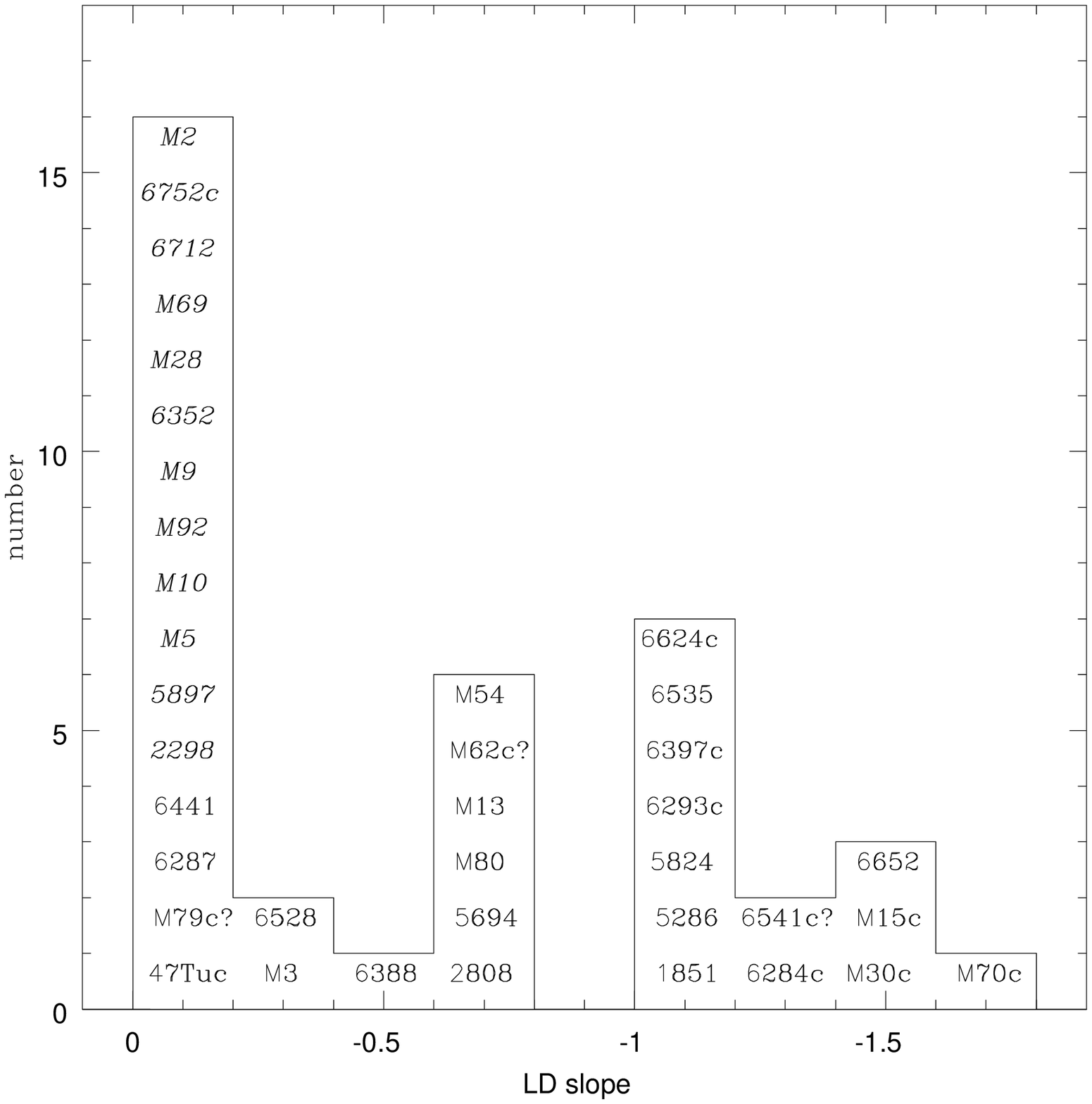,width=10cm,angle=0}}
  \figcaption{Histogram for luminosity density central logarithmic
  slopes. Cluster names are coded as in previous figure. Clusters in
  italics are those for which deprojection cannot be achieved due to
  diverging density profile near the center.}
\end{figure}

We need to compare the slope distributions with theoretical models for
globular clusters, particularly for those clusters with non-zero
slopes. As discussed in Section~1.2 there have been two mechanisms
explored for producing cusps in these systems: core-collapse and the
presence of an intermediate mass black hole in the center of the
cluster. The range of \mbox{3-dimensional} density slopes is narrower
for black hole than for core-collapse models, but they both center
around the same number $\sim-1.65$. However, only the clusters with
the steepest profiles in our sample fall in this range. In the case of
core-collapse the slope depends on the mass of the stars used to
construct the profile, so this could extend the range toward shallower
slopes. Another factor is the time dependence of the core-collapse
model when they go through the gravothermal oscillations. According to
Fokker-Planck simulations, a star cluster will spend a considerable
amount of time in between successive collapses, where the light
profile resembles a King model with a flat core. Unfortunately, these
models do not give enough details about the slope of the density
profile or the time spent on intermediate stages, so it is difficult
to say if the slope of our 'weak' cusp clusters are consistent with
this picture or if we need to invoke a new mechanism to explain this
shallower but non-zero slopes. We note that \citet{dul97} model M15 as
an intermediate stage of core-collapse. Since M15 has one of the
steepest profiles in our sample, then it appears that even invoking
this phase, it is unlikely to reproduce the full range that we find.

\begin{figure*}[t]
\centerline{\psfig{file=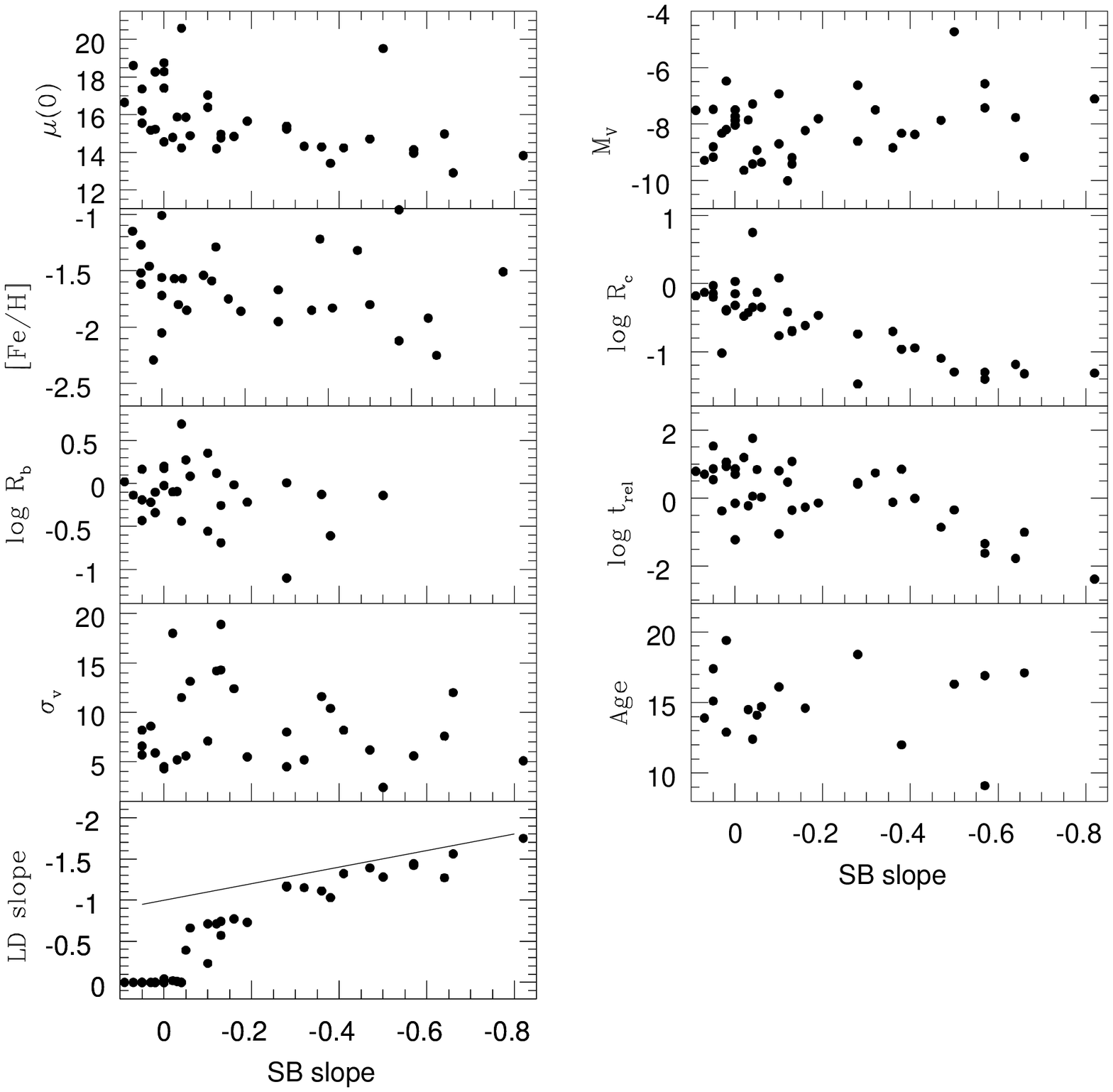,width=17.5cm,angle=0}}
  \caption{Surface brightness central logarithmic slope versus central
  surface brightness, absolute total V magnitude, metallicity,
  logarithmic core radius (in parsecs), logarithmic break radius (in
  parsecs), half light relaxation time, velocity dispersion,
  logarithmic age and luminosity density slope (the solid line
  represents `LD slope = SB slope + 1'). The distances to the clusters
  were obtained from Harris' catalog. There is a trend between central
  surface brightness and slope (with one obvious outlier). There is
  also a trend with core radius and relaxation time.}
\end{figure*}

\begin{figure*}[t]
\centerline{\psfig{file=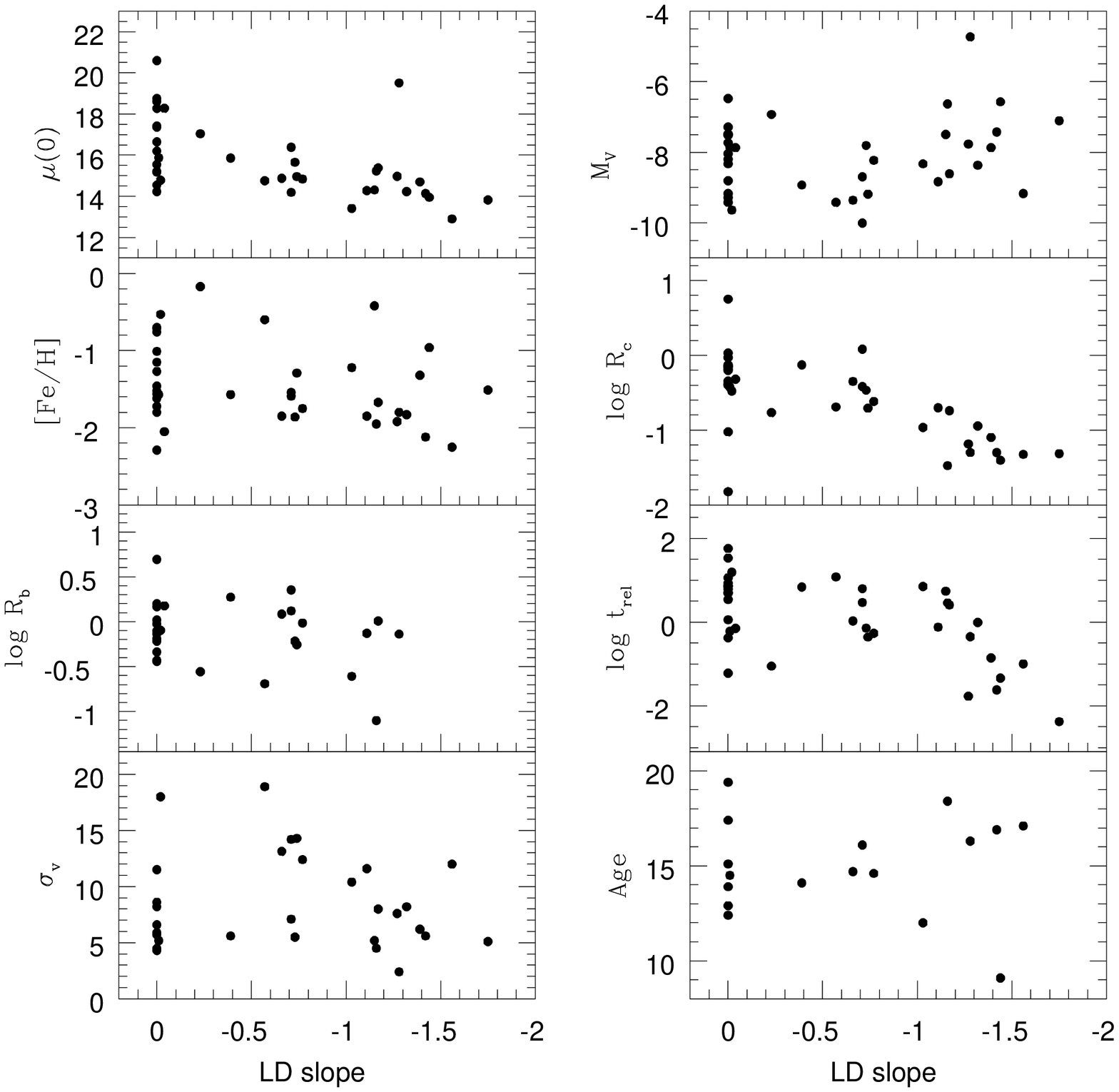,width=17.5cm,angle=0}}
  \caption{Luminosity density central slope versus central surface
  brightness, absolute total V magnitude, metallicity, logarithmic
  core radius, logarithmic break radius, half light relaxation time,
  velocity dispersion and logarithmic age.}
\end{figure*}

An alternative explanation for the existence of intermediate slopes is
presented by \citet{bau04}. They perform detailed numerical
simulations of clusters containing an intermediate-mass black hole in
their center. Their results show that the surface brightness profile
after a Hubble time shows a shallow cusp with slopes around $-0.25$,
and clearly distinguishable from zero. There are at least 8 objects in
our sample that fall into this category, but without complementary
kinematical measurements this hypothesis cannot be confirmed.

We plot logarithmic SB and LF central slopes against a variety of
global properties of clusters taken from Harris' catalog or measured
in this work. Figures~11 and 12 show these plots for both central
slope values versus central surface brightness, total V magnitude,
metallicity, logarithmic physical core radius, logarithmic physical
break radius, logarithmic half-light relaxation time, velocity
dispersion and age. Fig~11 also shows the relation between SB slope
and LF slope. We observe some global trends. As it is to be expected,
the clusters with steep profiles tend to have brighter central surface
brightness values, although the very sparse cluster NGC~6535 is an
outlier. There is an indication that objects with steeper cusps are
found in smaller objects (i.e. higher total magnitude); this trend is
more clear for luminosity density slopes. Metallicity measurements do
not appear to show any trend. The same is true for galactocentric
distance, except that the objects with steeper cusps are all close to
the center of the galaxy, but given the size of our sample this might
just be a small number effect. Half-light relaxation time seems to be
shorter for the steep cases. As it is to be expected, the core radius
is smaller for clusters with steep profiles, while the break radius
shows no correlation with slopes. Velocity dispersion and age show no
correlation with slopes. Finally, the relation between surface
brightness and luminosity density slope is not linear, as expected,
and is similar to that observed for galaxies \citep{geb96}.

The measured values for outer slopes range from $-1.0$ to $-2.5$ for
the clusters in the sample. When we plot these outer slopes values
versus global properties, and in particular versus either central SB
slope or concentration, we find no correlations. So as far as this
sample goes, we cannot distinguish between King-type or core-collapse
objects from the outer slope of the profiles. This is illustrated on
Fig~13 where we overplot all the observed profiles, scaled in surface
brightness and either their break radii (when they exist) or core
radii (for the others). The profiles are color coded according to the
classification given above for flat cores, weak cusps and steep
cusps. It can be observed that although the different groups can be
separated in the inner region, they do not seem to split into groups
in the outer region. This figure confirms once again that the profiles
cannot be clearly divided into flat cores and steep cusps, but that
they span a continuous range of central profiles.

\subsection{Individual objects}

\subsubsection{NGC~6397}

NGC~6397 is a peculiar object because it has always been considered to
be in core collapse due to its steep inner profile, but unlike other
objects considered to be in core-collapse, this one shows a sizable
core. \citet{lug95} report measuring a 4-10\arcsec\ core. Our
measurement for the break radius for this cluster is 2.1\arcsec. We
fit a power-law plus core function for the central region of the
profile and we find that the fit with a 4.5 \arcsec\ core radius is a
good fit, but only for the central 10\arcsec. It could be the case of
a partially resolved core. In previous studies the inner slope is
measured in a radial range extending well beyond the measured core
radius (as far as 100\arcsec). We measure inner slopes at the central
few arcseconds for all objects in our sample, therefore our slope
value for this object is much shallower than previous
measurements. Our $-0.37$ central slope value places this object in the
weak cusp category.

\vspace{20pt}
\subsubsection{NGC~6535}

NGC~6535 contains very few stars, therefore the image has low signal
and the measured profile looks very noisy. We decided to include it in
the sample because despite having so few stars, it shows a very steep
central surface brightness profile. The photometric data shown in
Trager's catalog for this cluster shows an important deviation
($\sim0.8$ mag) with respect to the Chebychev polynomial fit between
2\arcsec\ and 15\arcsec, where the photometric points are brighter than
the polynomial fit. PSF effects might have been responsible for
missing a cusp in this measurements.

\begin{figure}[t]
\centerline{\psfig{file=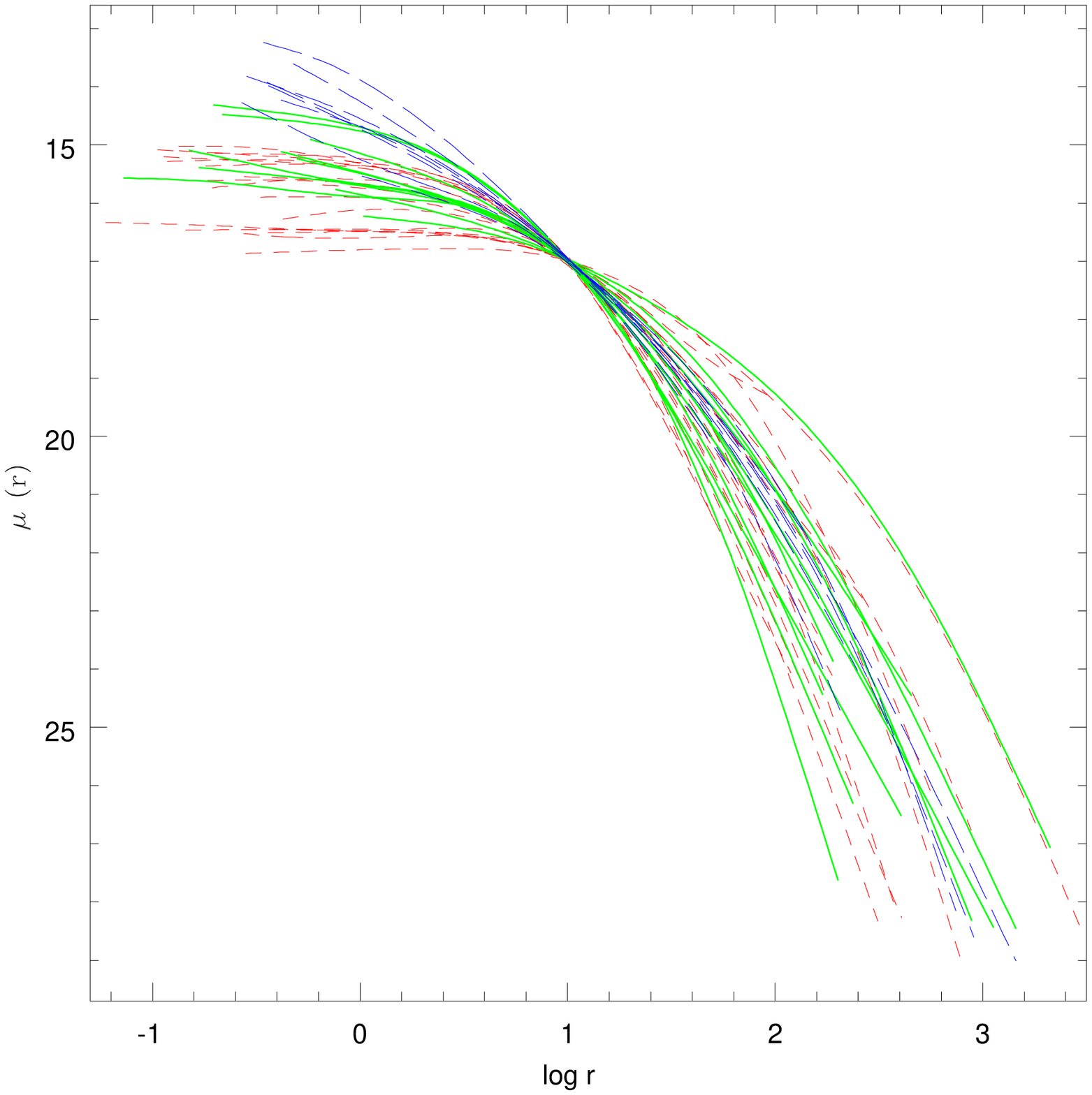,width=10cm,angle=0}}
  \figcaption{Surface brightness profiles for the entire sample. The
  profiles are normalized to a common point, therefore the units in
  the axis are arbitrary. Profiles are color coded according to their
  central slopes. Flat cores are shown in red (dashed lines), shallow
  cusps are shown in green (solid lines), and steep cusps are shown in
  blue (long dashed lines).}
\end{figure}

\begin{figure}[t]
\centerline{\psfig{file=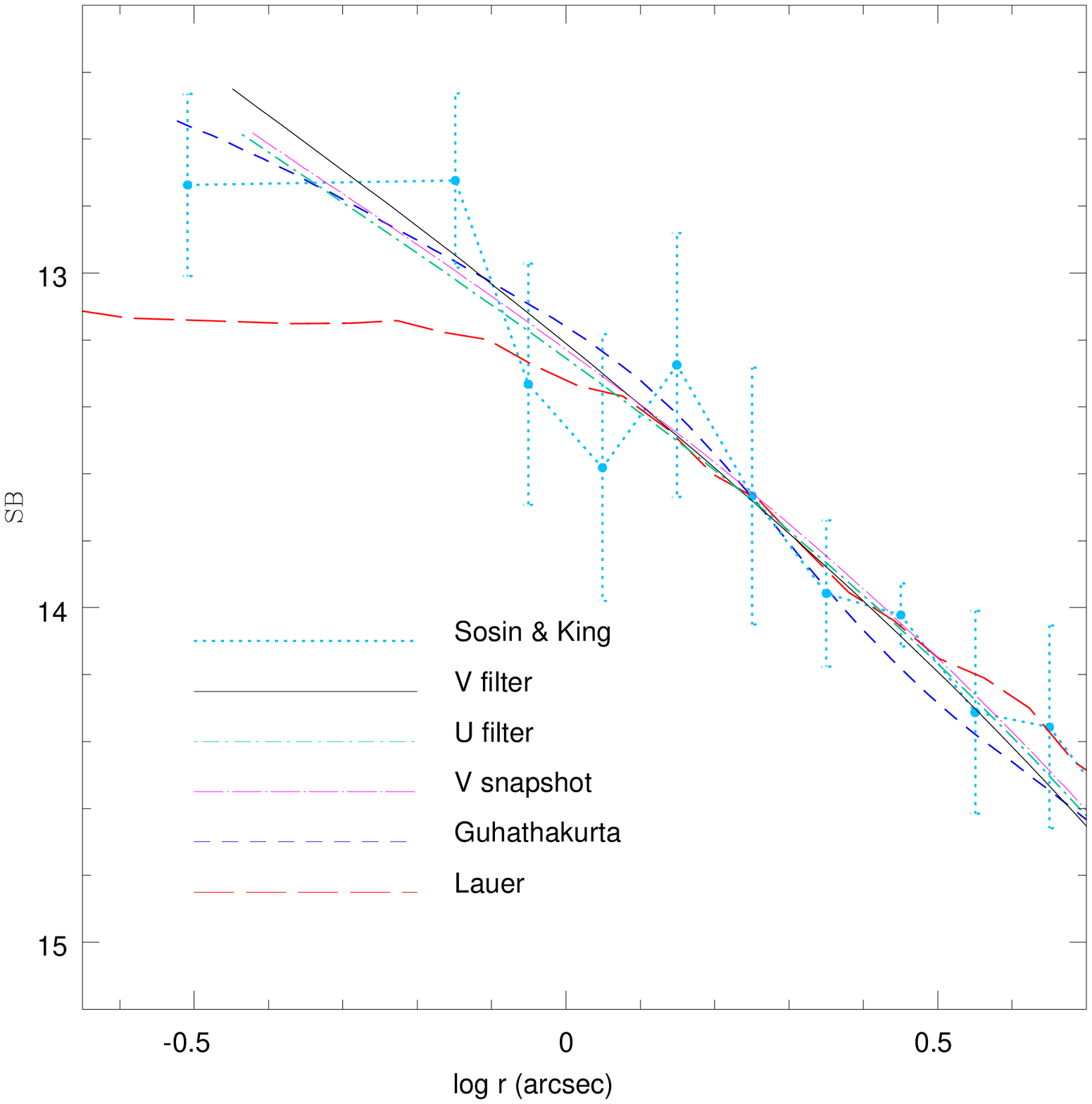,width=10cm,angle=0}}
  \caption{Surface brightness profiles for the central region of
  M15. Previously obtained profiles obtained by Guhathakurta (dashed
  dark blue), Sosin \& King (dotted light blue) and Lauer et al. (long
  dashed red) are plotted with our results from various images: long
  exposure V image (solid black), long exposure U image (dotted 
  -dashed green) and short exposure V (medium dashed magenta).}  
\end{figure}

\subsubsection{NGC~6652}

NGC~6652 is not considered to be in core collapse, but it shows a very
concentrated profile in our measurements. \citet{tra95} report a
4\arcsec\ core for this object. Our power-law plus core fit finds a
1.15\arcsec\ core and it is consistent with the photometry within the
error bars. This could be another case of a partially resolved
core. The central slope from the smooth profile is $-0.57$.

\subsubsection{NGC~6752}

NGC~6752 has been subject to a number of studies. This is the only
cluster in our sample for which we only analyzed the PC chip, without
including analysis of the WF chips. \citet{lug95} analyzed a
ground-based U-band image of the cluster and conclude that the surface
brightness profile does not present a core-collapse
morphology. \citet{fer03} constructed a surface density profile for
this cluster based on star counts. They fit the central region with
two separated King models, which they interpret as the cluster being
in post-core-collapse bounce. Our results indicate a flat core with a
slope near zero for both the surface brightness and luminosity density
profiles. Our difference from Ferraro et al. is likely due to noise in
the star counts that they use.

\subsubsection{M15}

There are a variety of WFPC2 images available for M15. For this reason
we applied the exact same procedure to each of them in order to test
the reliability of the profiles. We have a high signal-to-noise F555
image, a F336 image and a snapshot (60 sec) F555 image. In Fig~14 we
show our results for the inner part of the cluster, where we compare
them with previously obtained profiles by \citet{lau91}, \citet{gua96}
and \citet{sos97a}. Lauer et al's analysis used a WFPC1 image, where
they subtracted stars and measured the background starlight. Sosin \&
King's curve comes from star counts in a narrow magnitude range and
does not have any kind of smoothing applied to it, which is the reason
why it looks much noisier than the other curves. Guhathakurtha et al's
curve comes from corrected star counts and includes smoothing. All
three curves have an arbitrary vertical scaling. It can be seen that
the profiles are consistent in shape trough this radial range (inner 5
arcseconds), with the exception of Lauer's profile, which appears flat
toward the center. The center we measure is within 0.1 from that
obtained by both Guhathakurta and Sosin \& King, so we are confident
that center estimation is not a problem for this highly concentrated
object.

When measuring logarithmic inner slopes, the choice of the radial
extent used for the slope measurement is crucial. Sosin \& King
measure a $-0.7\pm0.5$ logarithmic slope by fitting a power-law over a
large radial extent between 0.3\arcsec\ and 10\arcsec. Guhathakurta et
al. report a slope of $-0.82\pm0.12$, again by fitting a power-law
between 0.3\arcsec\ and 6\arcsec; this power-law fits the star counts
near 6\arcsec but it is steeper than the points in the inner
0.5\arcsec. We measure the slope only for the innermost points
($<0.5$\arcsec) were it is a constant and get a value of
$-0.62\pm0.06$. If the same procedure is applied to Guhathakurta et
al.'s profile, we get a shallower slope of $-0.46$.

\section{Summary}

We obtain central surface brightness profiles for 38 galactic globular
clusters from $\it {HST}$/WFPC2 images in various filters. Generally,
we obtain reliable profiles into 0.5\arcsec. Based on extensive
simulations, we conclude that measuring integrated light with a robust
statistical estimator is superior for estimation of the profiles as
opposed to star counts when high signal to noise images are
available. Profiles obtained from images taken with different filters
are consistent and all are normalized to V-band by comparing to
profiles from ground-based data.

When compared with previous ground based measurements, our profiles
show different shapes for the inner regions. Most central surface
brightness measured are brighter than previously reported with values
up to two magnitudes brighter. The main reason for this difference is
the increased spatial resolution of $\it {HST}$, but also because we
use a non-parametric estimate as opposed to the traditional King model
fits. The full distribution of central slopes is not consistent with
simple isothermal cores. About half of our sample have a slope
distribution consistent with King models (i.e. flat core) and our
measurement uncertainties. The remaining 50\%, however, have a
distribution of SB logarithmic slopes that are fairly uniformly
distributed from $-0.2$ to $-0.8$. Our direct deprojection of the SB
profiles produces similar results for the luminosity density. About
half of the sample have luminosity density logarithmic slopes that
range from $-0.4$ to $-1.7$.

We find it challenging to explain these slope distributions when we
compare our results to existing dynamical models for globular
clusters, such as core-collapse or those that include a central black
hole. Both core-collapse and analytical black hole models predict
luminosity density slopes around $-1.6$. Core-collapse models can
accommodate the cases of intermediate slopes ($-0.2$ to $-0.5$ in SB,
and $-0.2$ to $-1.3$ in luminosity density) if we catch the clusters
at the appropriate time, and it seems unlikely to find them in the
high fraction that we measure. Recent numerical modeling for clusters
containing black holes \citep{bau04} might be able to explain some of
the intermediate slope cases.

Tables including our photometric measurements and fits can be found at:

www.as.utexas.edu/$\sim$eva/data.html.

\acknowledgements

This work is based in part on work supported by the Texas Advanced
Research Program under grant 003658--0243--2001. We acknowledge the
grant under HST-AR-09542 awarded by the Space Telescope Science
Institute, which is operated by the Association of the Universities
for Research in Astronomy, Inc., for NASA under contract NAS
5--26555. We also acknowledge the technical support from the Canadian
Astronomy Data Centre, which is operated by the Herzberg Institute of
Astrophysics, National Research Council of Canada. Finally, we
acknowledge the support by CONACYT.

\clearpage

\tabletypesize{\large}
\begin{deluxetable*}{lccrlcrc}
\tabletypesize{\large}
\tablecaption{\label{tbl1}Sample.}
\tablehead{
\colhead{NGC} & 
\colhead{other name}   & 
\colhead{filter} &
\colhead{exp. time} &
\colhead{image name} &
\colhead{$\alpha$ center} &
\colhead{$\delta$ center} & 
}
\startdata
104 & 47Tuc & F555 & 723 & u5470112b & 00:24:05.47 & -72:04:52.16  \\
1851 &\nodata  & F439 & 1200 & u2va0103b & 05:14:06.95 & -40:02:44.61 \\
1904 & M79 & F555 & 306 & u3ki0201b & 05:24:11.03 & -24:31:29.50 \\
2298 &\nodata& F814 & 905 &u3kt010gb  & 06:48:59.44 & -36:00:19.52  \\
2808 &\nodata& F555 & 314 & u4fp0105b & 09:12:03.09 & -64:51:48.96 \\
5272 & M3 & F555 & 1260 & u4r00101b & 13:42:11.33 &  28:22:37.81 \\
5286 &\nodata& F555 & 530 & u3um0201b & 13:46:26.73 & -51:22:28.77  \\
5694 &\nodata& F555 & 310 & u2y70105b & 14:39:36.29 & -26:32:20.19 \\
5824 &\nodata & F555& 320 & u2y70205b & 15:03:58.63 & -33:04:05.59 \\
5897 &\nodata& F555 & 608 & u3kt0204b & 15:17:24.50 & -21:00:37.00*  \\
5904 & M5 & F336 & 1200 & u3ki0302b & 15:18:33.36 & 02:04:55.19 \\
6093 & M80 & F675 & 780 & u3mu0104b & 16:17:02.48 & -22:58:33.18 \\
6205 & M13 & F555 & 2056 & u5bt0104b & 16:41:41.05 & 36:27:36.19 \\
6254 & M10 & F336 & 1500 & u3ki0102b & 16:57:08.9 & -09:05:58.0*  \\
6266 & M62 & F555 & 562 & u67e0209b & 17:01:12.96 & -30:06:46.20 \\
6284 &\nodata& F555 & 164 & u2xx0302b & 17:04:28.51 & -24:45:53.54 \\
6287 &\nodata& F555 & 3160 & u37a0106b & 17:05:09.13 & -22:42:30.14 \\
6293 &\nodata& F555 & 202 & u2xx0202b & 17:10:10.31 & -26:34:57.77 \\
6341 & M92 & F555 & 428  & u2z50109b & 17:17:07.34 & 43:08:10.08 \\
6333 & M9 & F555 & 2105 & u28q030lb & 17:19:11.26 & -18:30:57.41 \\
6352 &\nodata& F555 & 100 & u2kl0205b & 17:25:29.50 & -48:25.19.65 \\
6388 &\nodata& F336 & 1060 & u63t0301b & 17:36:17.18 & -44:44:07.83 \\
6397 &\nodata& F555 & 249 & u33r010kb & 17:40:41.57 & -53:40:26.03 \\
6441 &\nodata& F336 & 1060 & u63t0401b & 17:50:12.91 & -37:03:06.67 \\
6535 &\nodata& F555 & 1128 & u3kt040gb & 18:03:50.66 & -00:17:53.03 \\
6528 &\nodata& F555 & 814 & u61v0101b & 18:04:49.64 & -30:03:22.55 \\
6541 &\nodata& F555 & 596  & u28q050hb & 18:08:02.66 & -43:42:52.92 \\
6624 &\nodata& F555 & 1478  & u28q0604b & 18:23:40.22 & -30:21:41.32 \\
6626 & M28 & F555 & 1128 & u3kt050gb & 18:24:32.81 & -24:52:11.20 \\
6637 & M69 & F555 & 1690 & u28q0704b & 18:31:23.17 & -32.20:54.59 \\
6652 &\nodata& F555 & 1989  & u3m8010ib & 18:35:45.64 & -32:59:26.99 \\
6681 & M70 & F555 & 100 & u24s0103t & 18:43:12.83 & -32:17.33.38 \\ 
6712 &\nodata& F814 & 120 & u2of0205t & 18:53:04.30 & -08:42:22.0*  \\
6715 & M54 & F555 & 1850 & u37ga40cb & 18:55:03.29 & -30:28:46.10 \\
6752 &\nodata& F555 & 5246 & u2hO010cb & 19:10:52.237 & -59:59:03.81  \\
7078 & M15 & F555 & 400 & u2hr0102b & 21:29:58.40 & 12:10:00.26 \\
7089 & M2 & F555 & 106 & u67e0303b & 21:33:27.00 & -00:49:25.71 \\
7099 & M30 & F555 & 1192 & u5fw010nb & 21:40:22.16 & -23:10:47.64 \\ 
\enddata
\end{deluxetable*}

\clearpage
\begin{deluxetable*}{lccrrccccc}
\tabletypesize{\scriptsize}
\tablecaption{\label{tbl2}Measured parameters}
\tablehead{
\colhead{NGC} & 
\colhead{other name} &
\colhead{$\mu_V(0)$}   & 
\colhead{$r_c$}   & 
\colhead{$r_b$}   & 
\colhead{SB slope} &
\colhead{error   }&
\colhead{LD slope} &
\colhead{error   } &\\
\colhead{number}&
\colhead{} &
\colhead{(mag/arcsec$^2$)}   & 
\colhead{(arcsec)}   & 
\colhead{(arcsec)}   & 
\colhead{logarithmic} &
\colhead{}&
\colhead{logarithmic} &
\colhead{}&
}
\startdata
104 & 47Tuc & 14.35 &20.9 &16.4 & 0.00 & 0.04 & 0.11 & 0.15 \\
1851 &\nodata&  13.30 &2.0 &4.6 & --0.38 & 0.11 & --1.03 & 0.11 \\
1904 & M79 & 15.67 &5.6 &14.8 & --0.03 & 0.07 & --0.01 & 0.39 \\
2298 &\nodata& 18.72 &16.3 &17.4 & 0.00 & 0.07 & $\it0.00$ &\nodata  \\
2808 &\nodata& 14.89 &12.4 &36.1 & --0.06 & 0.07 & --0.66 & 0.54 \\
5272 & M3 & 15.72 &14.6 &46.9 & --0.05 & 0.10 &--0.39 & 0.45 \\
5286 &\nodata& 15.19 &4.2 &25.1 & --0.28 & 0.11  & --1.17 & 0.30 \\
5694 &\nodata& 15.62 &2.2 &2.6 &  --0.19 & 0.11 & --0.73 & 0.41 \\
5824 &\nodata& 14.17 &1.4 &4.0 & --0.36 & 0.16 &  --1.11 & 0.36 \\
5897 &\nodata& 20.47 &84.9 &119.0 &  --0.04 & 0.03 &$\it0.00$ &\nodata \\
5904 & M5 & 16.13 & 25.7 & 18.1 &  0.05 & 0.07 &$\it0.00$ &\nodata \\
6093 & M80 & 14.56 &4.5 &6.1&  --0.16 & 0.07 & --0.77 & 0.28 \\
6205 & M13 & 16.41 &34.4 &79.4 & --0.10 & 0.15 & --0.71 &0.32 \\
6254 & M10 & 17.68 &43.4 &22.4 & 0.05 & 0.07 &$\it0.00$ &\nodata \\
6266 & M62 & 14.78 &6.6 &13.8 &  --0.13 & 0.08 & --0.74 & 0.40 \\
6284 &\nodata& 14.66 & 1.1 &\nodata & --0.55 & 0.14 & --1.39 & 0.19 \\
6287 &\nodata& 18.32 & 11.3 & 34.4 & 0.00 & 0.07 & --0.04 &0.30 \\
6293 &\nodata& 14.43 & 1.0 &\nodata &  --0.67 & 0.08 & --1.27 & 0.18 \\
6341 & M92 & 15.29 &11.0 &17.15 & --0.01 & 0.04 &$\it0.00$ &\nodata \\
6333 & M9 & 17.01 &19.1 &41.8 & 0.00 & 0.13 & $\it0.00$ &\nodata \\
6352 &\nodata& 18.31 &23.2 &24.0 & 0.02 & 0.17 &$\it0.00$ &\nodata  \\
6388 &\nodata& 14.68 &4.4 &5.0 & --0.13 & 0.07 & --0.57 & 0.21 \\
6397 &\nodata& 15.29 &3.7 &2.7 & --0.37 & 0.11 & --1.16 & 0.20 \\
6441 &\nodata& 14.76 &5.8 &12.6 & --0.02 & 0.12 & --0.02 & 0.35 \\
6535 &\nodata& 19.35 &1.7 &21.2 & --0.50 & 0.18 & --1.28 & 0.38 \\
6528 &\nodata& 16.56 &3.9 &6.7 & --0.10 & 0.14 & --0.23 & 0.29 \\
6541 &\nodata& 14.38 &2.0 &\nodata & --0.41 & 0.09 & --1.32 & 0.22 \\
6624 &\nodata& 14.35 &1.7 &4.28 & --0.32 & 0.16 & --1.15 & 0.31 \\
6626 & M28 & 15.55 &9.8 &8.9& 0.03 & 0.05 &$\it0.00$ &\nodata \\
6637 & M69 & 16.71 &16.4 &49.5 & 0.09 & 0.13 &$\it0.00$ &\nodata \\
6652 &\nodata& 13.93 &1.2 &0.7 &  --0.57 & 0.12 & --1.44 & 0.20 \\
6681 & M70 & 13.68 &1.1 &\nodata &  --0.82 & 0.09 & --1.75 & 0.10 \\ 
6712 &\nodata& 18.57 &37.3 &68.6 & 0.02 & 0.05 &$\it0.00$ &\nodata \\
6715 & M54 & 14.12 &3.2 & 8.2& --0.12 & 0.07 & --0.71 & 0.35 \\
6752 &\nodata& 14.56 & 6.53 & 3.2 & --0.03 & 0.15 &$\it0.00$ &\nodata \\
7078 & M15 & 12.45 &0.98 &\nodata & --0.66 & 0.11 & --1.56 & 0.22 \\
7089 & M2 & 15.19 &12.9& 20.8 & 0.05 & 0.11 &$\it0.00$ &\nodata \\
7099 & M30 & 14.22 &1.6 &\nodata & --0.57 & 0.11 & --1.42 & 0.18 \\ 
\enddata

\tablecomments{col 1-2 are NGC and other names, col 3 is central
surface brightness in V, col 4 is core radius, col 5 is break
radius (as defined on Section~3.2), col 6-7 are logarithmic central
surface brightness slope and uncertainty, col 8-9 are logarithmic
central luminosity density slope and uncertainty.}

\end{deluxetable*}

\end{document}